\documentclass[reprint, amsmath,amssymb, aps, groupedaddress, a4paper]{revtex4-2}

\usepackage{amsmath}
\usepackage{placeins}

\usepackage{xcolor}
\usepackage{siunitx}
\usepackage[version=4]{mhchem}

\usepackage[a4paper,margin=2cm]{geometry}
\usepackage{graphicx}
\usepackage{dcolumn}
\usepackage{bm}
\usepackage{hyperref}
\usepackage[mathlines]{lineno}

\begin{document}

\preprint{APS/123-QED}

\title{Visual and quantitative analysis of the trapping volume\\ in dielectrophoresis of nanoparticles}

\author{Siarhei Zavatski}
 \email{siarhei.zavatski@epfl.ch}
\author{Olivier J.F. Martin}%
 \email{olivier.martin@epfl.ch}
\affiliation{%
 Nanophotonics and Metrology Laboratory (NAM),\\
 Swiss Federal Institute of Technology Lausanne (EPFL),\\
 Lausanne 1015, Switzerland
}%

\date{June 19, 2024}

\begin{abstract}
Nanoparticle manipulations require a careful analysis of the forces at play. Unfortunately, traditional force measurement techniques based on the particle velocity do not provide a sufficient resolution, while balancing approaches involving counteracting forces are often cumbersome. Here, we demonstrate that a nanoparticle dielectrophoretic response can be quantitatively studied by a straightforward visual delineation of the dielectrophoretic trapping volume. We reveal this volume by detecting the width of the region depleted of gold nanoparticles by the dielectrophoretic force. Comparison of the measured widths for various nanoparticle sizes with numerical simulations obtained by solving the particle–conservation equation shows excellent agreement, thus providing access to the particles physical properties, such as polarizability and size. These findings can be further extended to investigate various types of nano–objects $-$ including bio– and molecular aggregates $-$ and offer a robust characterization tool that can enhance the control of matter at the nanoscale. 
\end{abstract}

\maketitle


Electrokinetic effects enable the precise and long–range control of the position of numerous micro- and nanoscale species. As such, they have tremendous potential for both fundamental~\cite{ramos_ac_1998,schoch_transport_2008,squires_microfluidics_2005,riccardi_electromagnetic_2023} and applied research~\cite{yuan_electrokinetic_2007,llorente_applications_2014,mir_electrokinetic_2011}. For example, dielectrophoresis (DEP) can renovate the field of separation techniques ~\cite{pethig_dielectrophoresis_2010, pethig_dielectrophoresis_2017}. Indeed, there is a solid body of research that features the successful utilization of the DEP force for transport~\cite{camacho-alanis_transitioning_2012}, trapping \cite{hughes_dielectrophoretic_1998,holzel_trapping_2005}, separation \cite{liu_orders--magnitude_2021, regtmeier_dielectrophoretic_2007, yunus_continuous_2013}, and concentration \cite{liao_nanoscale_2012, lalonde_isolation_2015, damico_isolation_2017, nguyen_impedance_2018} of different inorganic and biological substances. However, a reliable DEP experiment requires a valid experimental estimate of the DEP force, which is usually not straightforward. There is no possibility to measure the DEP force directly and it is typically estimated indirectly, which is possible only as long as a precise theoretical model for DEP exists; unfortunately this may not always be the case, e.g. for submicron bioparticles \cite{pethig_limitations_2019, pethig_protein_2022,holzel_protein_2020,holzel_protein_2021,seyedi_protein_2018,heyden_dielectrophoresis_2020}. Therefore, developing new force measurement strategies is of fundamental interest for DEP research and its application in nanosciences.

Several approaches have been proposed to measure the DEP force \cite{hoettges_dielectrophoresis_2010}. The most common one relies on estimating the particle velocity from videos recorded on an optical microscope \cite{watarai_situ_1997,huang_differences_1992,kralj_continuous_2006,ai_transient_2009}. The DEP force can then be determined by solving the Langevin equation \cite{lemons_paul_1997,langevin_sur_1908}. However, a reliable force estimate obtained this way also requires the correct definition of all the other forces that may act on the particle during DEP. Furthermore, if the particles are unlabeled and in low concentration, this method is unsuitable for nanoscale particulates, simply because their observation in an optical microscope is challenging. Alternatively, the DEP force can be measured by a balancing approach that requires another counteracting force of known magnitude, such that the total force on the target object vanishes. For example, the counteracting force can be optical \cite{park_direct_2014,hong_quantitative_2010,jeon_dielectrophoretic_2017}, gravity \cite{imasato_measurement_2008}, drag \cite{liu_orders--magnitude_2021,lu_dielectrophoretic_2020,su_rapid_2013}, or thermal randomizing caused by the Brownian motion \cite{zavatski_protein_2023}. We recently used the latter with a gradient array of conductive electrodes to measure the DEP polarizability factors for three proteins \cite{zavatski_protein_2023}. Unfortunately, the proposed electrodes cannot be utilized to investigate a negative DEP force and the corresponding protein polarizability because their configuration does not provide clearly defined regions with minimum electric field gradient intensities, where the negative DEP trapping can be detected. Other strategies are also available to measure the DEP force, including measurements of the collection rate \cite{labeed_assessment_2003,labeed_differences_2006,markx_dielectrophoretic_1994,hubner_water_2003}, cross–over frequency \cite{hughes_manipulation_1998,green_manipulation_1997,schnelle_trapping_1996}, and levitation height \cite{jones_bubble_2008,kaler_dielectrophoretic_1990}. 

Here, we report a straightforward visual representation and quantitative estimate of a particle DEP response, which relies on revealing the DEP trapping volume. The key advantage of this technique is that it does not require special electrodes design or complicated experimental setups to gain a quantitative description of the particle movement. Rather, it can be applied to any DEP platform to reveal the interplay between different forces acting on the particle during the experiment. Besides, it may be applied to investigate any substance in both negative and positive DEP regimes, thus providing the frequency dependence of the DEP polarizability. Furthermore, it can be used to gain quantitative understanding of the temperature, pH, and conductivity dependencies of the DEP polarizability. All of this can be extremely useful for addressing fundamental challenges in DEP, such as the development and verification of new DEP models for the accurate ab initio simulations of the DEP response of bio–nanoparticles \cite{pethig_protein_2022,holzel_protein_2020,holzel_protein_2021,heyden_dielectrophoresis_2020}.

\begin{figure*}
  \centering\includegraphics{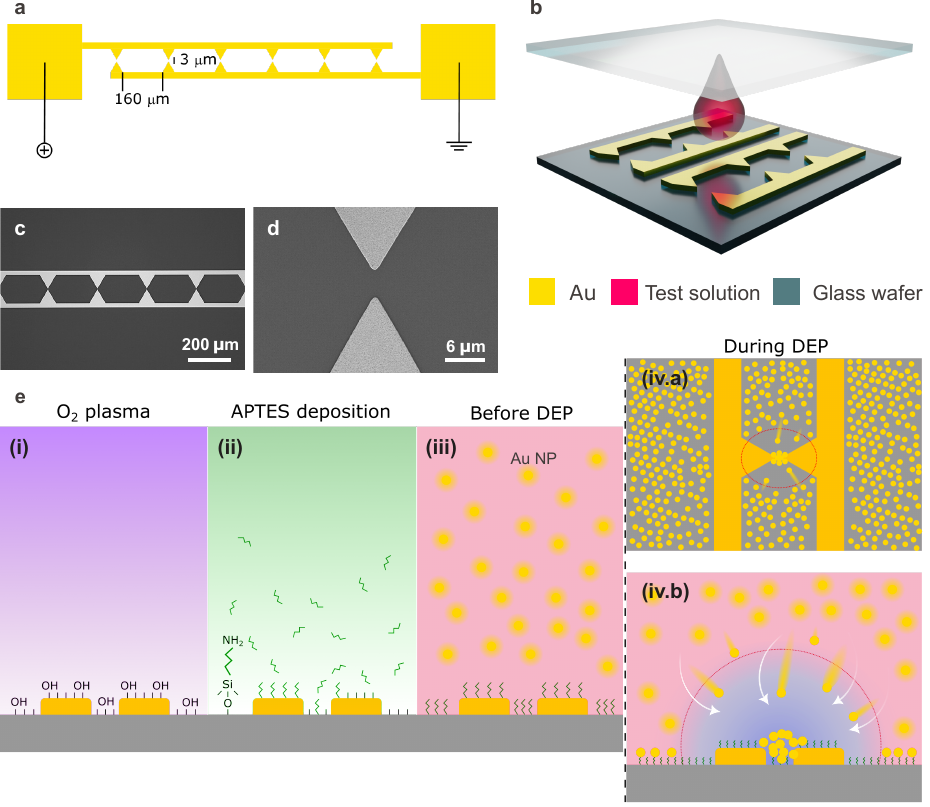}
  \caption{(a) DEP device design showing the unit cell for the sawtooth metal electrode array and (b) schematic representation of the microfluidic chamber utilized for the DEP experiments. (c) Optical microscope and (d) SEM images of a sawtooth metal electrode array (top view). (e) Schematic representation of the DEP device preparation (cross–sectional view) and working principle utilized to visualize the depletion region: (i) DEP device surface cleaning and hydro–oxidation by oxygen plasma treatment; (ii) gas phase (3–aminopropyl)triethoxysilane (APTES) deposition on top of the OH–rich DEP device surface; (iii) the experimental system before AC voltage application, after addition of Au nanoparticles and microfluidic chamber assembly; (iv.a) top and (iv.b) cross–sectional views of the experimental system during the DEP experiment. Au nanoparticles outside the depletion region indicated by the red circle attach to the primary amine (NH\textsubscript{2}–) of APTES molecules through a diffusion–limited process. Au nanoparticles inside the red circle region move towards and accumulate near the sawtooth electrode apexes. This produces two distinct areas on the surface, with high and low concentrations, which may be observed by dark-field microscopy.}
  \label{fgr:fig1}
\end{figure*}

The DEP platform utilized in this work is depicted in Figures \ref{fgr:fig1}a–d (see Materials and Methods in the Supporting Information for fabrication details). It consists of periodically repeated sawtooth gold electrode pairs on a glass substrate separated by a fixed gap of 3 $\mu$m. The lateral distance between sawtooth gaps is 250 $\mu$m to avoid any coupling between adjacent electrodes. The DEP is readily observed when these electrodes are immersed in an aqueous dispersion of nanoparticles and energized by an external electrical signal. The time-averaged DEP force acting on a nanoparticle in solution is defined as \cite{pethig_dielectrophoresis_2010,pethig_dielectrophoresis_2017}
\begin{equation}
  \bigl \langle\mathbf{F_{DEP}} \bigr \rangle=\pi R^3 \varepsilon_m \varepsilon_0 \operatorname{Re}\left[\frac{\varepsilon_p - \varepsilon_m}{\varepsilon_p+2\varepsilon_m}\right] \nabla |{\mathbf{E}}|^2 \label{eqn:eq1}
\end{equation}
where R is the particle radius, $ \varepsilon_p $ the dielectric constant of the particle, $ \varepsilon_m $ the medium dielectric constant, $ \varepsilon_0 $ the vacuum permittivity, and $ |{\mathbf{E}}| $ the amplitude of the electric field. The term in square brackets in Eq.~(\ref{eqn:eq1}) is the real part of the Clausius–Mossotti (CM) or DEP polarizability factor – the most critical and intricate parameter for the accurate description of DEP \cite{pethig_limitations_2019,pethig_protein_2022}. It does not only determine the direction of a particle movement in an inhomogeneous electric field but also influences the magnitude of the DEP force \cite{pethig_dielectrophoresis_2017}.

Our hypothesis to experimentally estimate the DEP parameters in Eq.~(\ref{eqn:eq1}) is that two distinct volumes must appear near the electrodes during a DEP trapping experiment, with respectively high and low concentrations of nanoparticles. The volume with a low concentration – also known as the depletion or trapping volume \cite{bakewell_dielectrophoresis_2006,cummings_dielectrophoresis_2003,cummings_streaming_2003,ding_concentration_2016} – is where DEP translates nanoparticles towards (positive DEP) or away from (negative DEP) the strongest electric field gradient. This translation occurs because the time-averaged DEP potential energy, $\bigl \langle U_{DEP}\bigr \rangle$, of nanoparticles inside the trapping volume, exceeds the thermal randomizing energy, $3k_BT/2$ \cite{ramos_ac_1998,washizu_molecular_1994}:
\begin{equation}
  \bigl \langle U_{DEP}\bigr \rangle = -\pi R^3 \varepsilon_m \varepsilon_0 \operatorname{Re}\left[\frac{\varepsilon_p - \varepsilon_m}{\varepsilon_p+2\varepsilon_m}\right] |{\mathbf{E}}|^2 \label{eqn:eq2}
\end{equation}
\begin{equation}
  \bigl \langle U_{DEP} \bigr \rangle > \frac{3}{2} k_BT \label{eqn:eq3}
\end{equation}
where $k_B$ is Boltzmann’s constant and $T$ the absolute temperature.

Figure~\ref{fgr:fig1}e sketches the measurement procedure of the DEP trapping volume. To test our hypothesis, we use Au nanoparticles of different radii, although the technique is applicable to any substance. The cross–section of the trapping volume is experimentally recorded by analyzing the dark–field scattering from Au nanoparticles immobilized on the DEP device surface. To provide an appropriate contrast between high and low (i.e., depleted by DEP) concentration regions, we also functionalize the device with APTES (Figure~\ref{fgr:fig1}e, steps (i)–(iii)). In the absence of DEP, APTES ensures a strong binding of nanoparticles to the surface, producing a uniform nanoparticle layer evidenced by a smooth background scattering intensity. This layer slowly builds up everywhere on the surface by the diffusion–limited motion of nanoparticles, Figure~\ref{fgr:fig1}e(iii). On the other hand, when the electric field is applied to the electrodes and DEP sets in, nanoparticles are rapidly moved by DEP from within the depletion region to the electrode apexes, preventing interaction with APTES. This leads to a local depletion of the number of nanoparticles adsorbed on the surface, which reduces the dark–field scattering intensity from this region, as illustrated in panels (iv.a) and (iv.b) in Figure~\ref{fgr:fig1}e. The scattering intensity is recorded and analyzed to obtain its spatial profile.

In general, the concentration of Au nanoparticles in DEP experiments evolves as the result of the interplay between nanoparticle drift and subsequent diffusion process caused by their DEP–induced redistribution in space. Assuming an ensemble of non–interacting nanoparticles, this concentration profile is given by the particle–conservation equation \cite{kittel_elementary_2004,bakewell_modelling_2011,loucaides_dielectrophoretic_2011}:
\begin{equation}
  \frac{\partial c}{\partial t} + \nabla \cdot \bigl(c \mathbf{u_{fluid}} + \mathbf{J_{T}}\bigr) = 0, \label{eqn:eq4}
\end{equation}
where $ c=nV_p $ is the volume fraction of particles (referred further as the concentration, for brevity) with particle number density $n$ and volume $V_p$, $\mathbf{u_{fluid}}$ is the velocity of the liquid medium, and $\mathbf{J_{T}}$ is the total flux consisting of the sum of the diffusion, $\mathbf{J_{D}}$, sedimentation, $\mathbf{J_{sedim}}$, and DEP fluxes, $\mathbf{J_{DEP}}$:
\begin{equation}
  \mathbf{J_{T}} = \mathbf{J_{D}} + \mathbf{J_{sedim}} + \mathbf{J_{DEP}}, \label{eqn:eq5}
\end{equation}
with
\begin{equation}
  \mathbf{J_{D}} = -D \nabla c, \label{eqn:eq6}
\end{equation}
\begin{equation}
  \mathbf{J_{sedim}} = \frac{c\mathbf{F_{sedim}}}{6 \pi \eta R}, \label{eqn:eq7}
\end{equation}
\begin{equation}
  \mathbf{J_{DEP}} = \frac{c\mathbf{F_{DEP}}}{6 \pi \eta R}, \label{eqn:eq8}
\end{equation}
where $D = k_BT/6 \pi \eta R$ is the diffusion coefficient for Au nanoparticles, $\eta $ the liquid viscosity, and  $\mathbf{F_{sedim}} = \left( \rho_m - \rho_p \right) V_p g $ is the sedimentation force with $\rho_m$ the medium and $\rho_p$ the particle densities, and gravitational acceleration $g$ \cite{ramos_ac_1998,castellanos_electrohydrodynamics_2003}. 

The solution of Eq. (\ref{eqn:eq4}) provides the spatial–temporal evolution of the nanoparticle concentration, which can be effectively compared with experimental results and used to quantitatively characterize the DEP response of a particle. However, obtaining this solution for specific experimental conditions is not straightforward and requires a careful definition of initial and boundary conditions \cite{bakewell_modelling_2011}.

In this work, we obtain quantitative information on DEP by comparing the size of the low dark–field intensity measured on the DEP device surface with numerical simulations obtained by solving Eq. (\ref{eqn:eq4}) assuming stationary conditions, such that the first term on the left–hand side vanishes. This simplification is possible because the experimental conditions are usually long enough to reach equilibrium between the DEP–induced transport and the diffusion of particles. The results obtained by A. Castellanos et al. also suggest that we can neglect the sedimentation flux defined in Eq. (\ref{eqn:eq7}) because the displacement caused by gravity and buoyancy for particles with a 25–75 nm radius in water is smaller than the displacements induced by DEP and thermal perturbations \cite{castellanos_electrohydrodynamics_2003}. Finally, our experimental conditions, including low buffer conductivity (16 $\mu$S/cm) and an optimized frequency of the applied electric field (3 MHz), suppress the bulk fluid movement upon DEP, and convection, $\mathbf{u_{fluid}}$, vanishes. As a result, the solution of Eq. (\ref{eqn:eq4}) for the nanoparticle concentration on the DEP device surface takes the following form \cite{ramos_ac_1998,bakewell_modelling_2011}:
\begin{equation}
  c_{surf} \left(x,y \right) = Aexp\left(-\frac{|\bigl \langle U_{DEP}\bigr \rangle|}{k_BT}\right), \label{eqn:eq9}
\end{equation}
where $A$ is an arbitrary integration constant. The exponential in Eq. (\ref{eqn:eq9}) echoes the condition introduced in Eq. (\ref{eqn:eq3}) and indicates that the boundaries between depleted and undepleted regions are smeared out for an ensemble of nanoparticles due to their random thermal perturbation.

\begin{figure*}
  \centering\includegraphics{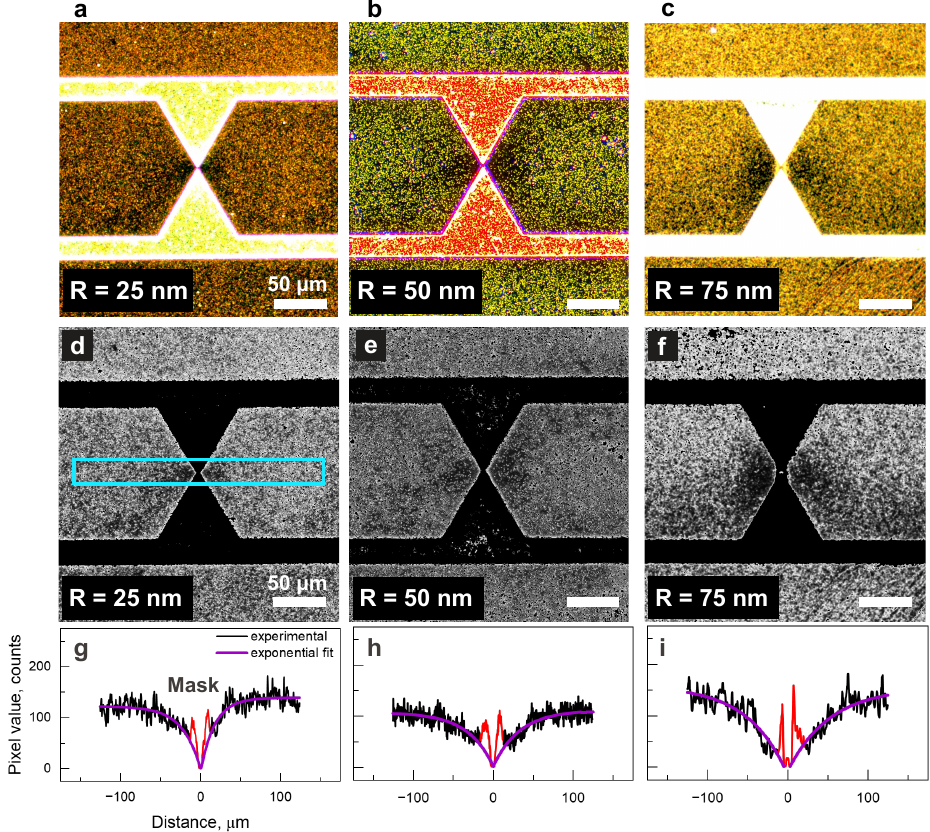}
  \caption{(a-c) Dark–field images acquired for Au nanoparticles with the radius of (a, d) 25 nm, (b, e) 50 nm, and (c, f) 75 nm after DEP at 15 V\textsubscript{p-p} and 3 MHz. (d-f) Dark–field scattering intensity profiles obtained by integrating within a rectangle shown in cyan in panel (a) (see text for details). An exponential fit, shown in purple, is obtained by masking the data near the electrode gap, where the saturated optical signal is incompletely subtracted. All scale bars are 50 $\mu$m.}
  \label{fgr:fig2}
\end{figure*}

It should also be noted that the argument of the exponential in Eq. (\ref{eqn:eq9}) is negative because we study the surface concentration of Au nanoparticles, which is assumed to be the inverse of the bulk concentration profile. This approximation is justified for the following reasons. In positive DEP experiments, the bulk particle concentration near electrodes increases progressively with time, reaching the steady–state maximum value in the strongest electric field gradient region (see Figure S1 in the Supporting Information). At the same time, as revealed in our simulations, the perpendicular component of the DEP force, $\bigl \langle\mathbf{F_{DEP}}(z)\bigr \rangle$, which is responsible for nanoparticle translation to and adsorption onto the surface, is much weaker compared to the lateral ones, $\bigl \langle\mathbf{F_{DEP}}(x)\bigr \rangle$ and $\bigl \langle\mathbf{F_{DEP}}(y)\bigr \rangle$ which induce nanoparticles movement parallel to the surface (see Figure S2 in the Supporting Information). Hence, we can assume that the DEP device surface is depleted in the trapping volume by roughly the same number of nanoparticles as accumulated in the bulk liquid just above the surface.

\begin{figure*}
  \centering\includegraphics{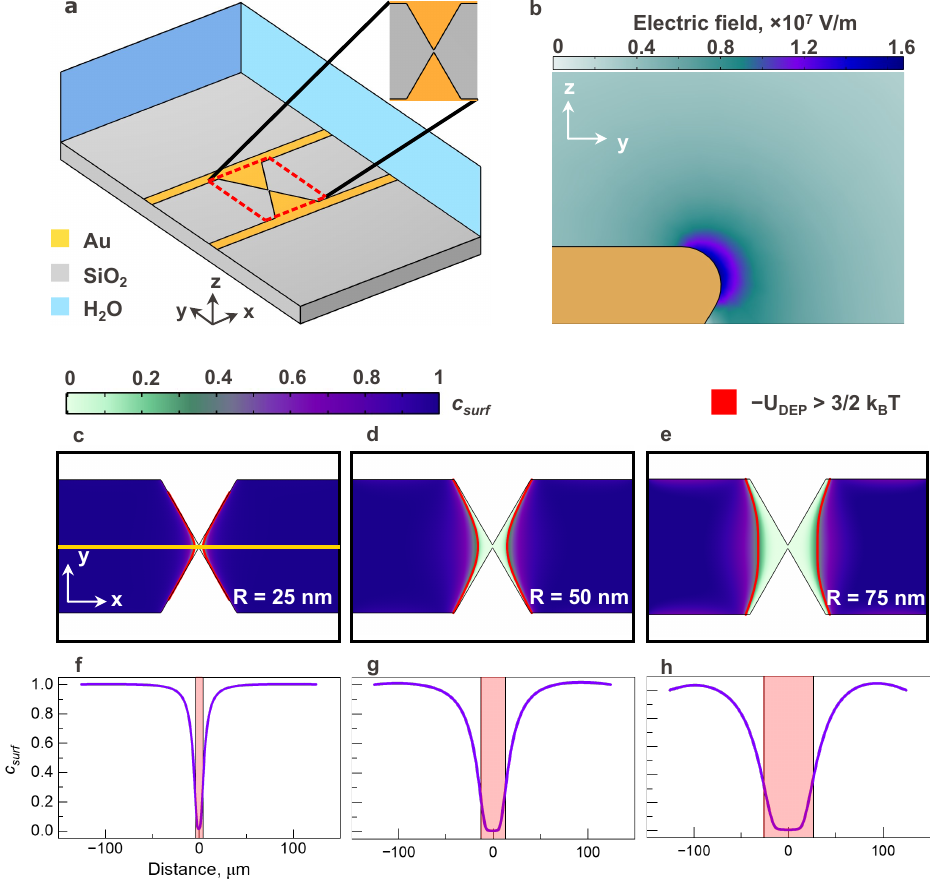}
  \caption{(a) 3D geometry of the DEP device used to simulate (b) the electric field intensity distribution near sawtooth metal electrodes. (c–h) 2D simulation results of the concentration distributions for (c, f) 25 nm, (d, g) 50 nm, and (e, h) 75 nm radius Au nanoparticles after applying a sinusoidal electric signal with 15 V\textsubscript{p-p} peak–to–peak voltage and 3 MHz frequency. The concentration distribution profiles in (f–h) were calculated along the yellow line crossing the middle of the gap between adjacent electrode pairs. The red contours in (c–e) and bands in (f–h) depict the area where the DEP potential energy is larger than the thermal diffusion energy, $\bigl \langle U_{DEP} \bigr \rangle > 3k_BT/2$.}
  \label{fgr:fig3}
\end{figure*}

Let us first support the proposed hypothesis and above analysis with experimental results. Figures \ref{fgr:fig2}a–c show the dark–field scattering images acquired for Au nanoparticles with different radii after DEP (15 V\textsubscript{p-p} and 3 MHz). The regions of high and low dark–field scattering intensities are well visible, with the lowest intensity near the electrode apexes and a progressive intensity increase as one moves away from them. At some distance from the electrode gap, the dark–field intensity reaches a certain magnitude and then remains constant, indicating that one has left the depletion region and entered undepleted space, where the exponential in Eq. (\ref{eqn:eq9}) becomes negligible. Figure \ref{fgr:fig2} also indicates that this transition is observed farther from the electrode gap for larger nanoparticles, which is also expected from Eq. (\ref{eqn:eq9}) because the DEP potential energy of nanoparticles has a cubic dependence on their radii. Hence, the obtained scattering profiles can be reliably attributed to the generation of the DEP depletion region.

Let us now compare experimental data with simulated concentration profiles. The 3D simulation domain and corresponding electric field intensity distribution near the sawtooth electrode apex are shown in Figure \ref{fgr:fig3}a and \ref{fgr:fig3}b (see Materials and Methods in the Supporting Information for additional simulation details). This model is based on the effectively fabricated geometry, as shown in Figures \ref{fgr:fig1}c,d. The geometrical parameters, including the radii of curvature utilized to simulate the electrode tip apex, were carefully determined using SEM and focused ion beam images \cite{zavatski_protein_2023}. A maximum electric field intensity of $1.72 \cdot 10^7$ V/m was calculated near the electrode apex for an applied peak–to–peak voltage of 15 V.

We utilize the electric field components $ \mathbf{E}(x) $ and $ \mathbf{E}(y) $ simulated in 3D to compute   in the plane of the DEP device surface, see Eq. (\ref{eqn:eq8}), and calculate the Au nanoparticle concentration distributions, $ c_{surf}(x, y) $ by solving Eq. (\ref{eqn:eq4}) in 2D (see Materials and Methods in the Supporting Information for additional simulation details that indicate that the same concentration profiles are observed when Eq. (\ref{eqn:eq4}) is solved in 3D). The obtained concentration profiles are shown in Figures \ref{fgr:fig3}c–h. Figures \ref{fgr:fig3}c–e show the spatial variation of the concentration near the electrodes, while Figures \ref{fgr:fig3}f–h depict the same concentration profiles along the yellow line in Figure \ref{fgr:fig3}c. These figures indicate a significant concentration variation near the electrodes, revealing the shape and size of the depletion regions, which are in a good agreement with the experimental scattering profiles shown in Figure \ref{fgr:fig2}a–i. The minimum of surface concentration is observed for all the studied nanoparticles in the middle of the gap between the electrodes. It gradually increases with the distance from the gap, approaching the high concentration limit. Besides, Figures \ref{fgr:fig3}c–h indicate that the depletion region is wider for larger Au nanoparticle, which is again in agreement with the scaling of $ \bigl \langle U_{DEP}\bigr \rangle $ defined by Eq. (\ref{eqn:eq2}).
\begin{table*}[htp!]
\caption{\label{tab:table1}%
Comparison of experimental and simulated DEP depletion region sizes, and corresponding particle radii calculated by Eq. (S2). All experimental values are obtained by analyzing dark–field scattering intensities from approx. 50 electrode pairs.}
\begin{ruledtabular}
\begin{tabular}{cccc}
 &\multicolumn{2}{c}{Depletion region size, $\mu$m}&\multicolumn{1}{c}{\textit{R}, nm}\\
 
\textrm{Au nominal radius, nm}&
\textrm{Experimental}&
\multicolumn{1}{c}{Simulated}&
\textrm{Calculated by Eq. (S2)}\\
\colrule
25 & 16.0 $\pm$ 4.4 & 9.7 & 30.8 $\pm$ 2.9\\
50 & 35.6 $\pm$ 8.8 & 36.0 & 48.5 $\pm$ 7.9\\
75 & 61.4 $\pm$ 12.4 & 63.7 & 71.1 $\pm$ 10.7\\
\end{tabular}
\end{ruledtabular}
\end{table*}

At this point, we should emphasize the importance of considering nanoparticle diffusion to simulate the depletion region size. This can be observed in Figures \ref{fgr:fig3}c–h, where the red contour depicts the spatial extension of the condition in Eq. (\ref{eqn:eq3}). The width of the depletion region defined by Eq. (\ref{eqn:eq3}) and calculated along the yellow line in Figure \ref{fgr:fig3}c, varies with the nanoparticle radius: 7.6 $\mu$m, 25.9 $\mu$m, and respectively 52.4 $\mu$m for 25 nm, 50 nm, and respectively 75 nm Au nanoparticles. It is noteworthy that the actual width of the surface concentration variation can be significantly larger than that obtained by balancing the thermal energy, especially for small particles sizes (see the concentration value at which the red band crosses the concentration profile for various Au nanoparticle radii in Figure \ref{fgr:fig3}f-h). These results indicate that – when investigating the DEP of a nanoparticle ensemble – the trapping volumes must be estimated by applying the laws of statistical physics.

\begin{figure}[hbp!]
  \centering\includegraphics{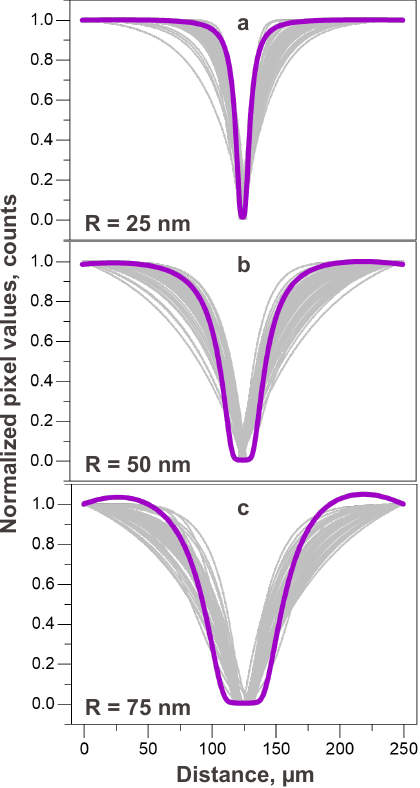}
  \caption{Quantitative analysis of the dark–field scattering intensity profiles acquired from approx. 50 different electrode pairs for (a) 25 nm, (b) 50 nm, and (c) 75 nm Au nanoparticles at 15 V\textsubscript{p-p} and 3 MHz. The gray lines represent the experimental exponential fits obtained with the procedure outlined in the main text and are similar to those depicted in Figure \ref{fgr:fig2}g–i. The purple lines correspond to the simulated concentration profiles shown in Figure \ref{fgr:fig3}f–h. Each experimental profile has been normalized between its minimum and its maximum.}
  \label{fgr:fig4}
\end{figure}

Let us now compare the obtained experimental scattering profiles with the simulated concentration distributions of Au nanoparticles. Figure \ref{fgr:fig4} shows the dark–field scattering intensity fits for approx. 50 various sawtooth microelectrode pairs. The gray curves represent the corresponding concentration profiles for Au nanoparticles with radii of 25 nm (Figure \ref{fgr:fig4}a), 50 nm (Figure \ref{fgr:fig4}b), and 75 nm (Figure \ref{fgr:fig4}c). The purple lines in Figure \ref{fgr:fig4} correspond to the simulated concentration profiles shown in Figure \ref{fgr:fig3}f–h. The average experimental sizes of the depletion region after DEP for various nanoparticles are shown in Table 1. These values are in very good agreement with the simulation results.

To demonstrate that the proposed approach can be effectively utilized for the quantitative characterization of the DEP response of different nanoscopic objects, we analyze the obtained concentration profiles and estimate the Au nanoparticle radius (see the Supporting Information for calculation procedure). The calculation results are summarized in the last column of Table 1. The agreement between the experimentally deduced radii and their nominal values is excellent. This approach is very general and can be used to determine any parameter in Eq. (S2), including the DEP polarizability factor.

Let us also note that the accuracy achieved here results from a careful optimization of the experimental conditions. At least two critical factors may lead to significant errors and must be carefully handled for the correct estimation of the depletion region (see the Supporting Information for further discussion). One relates to the first term in brackets in Eq. (\ref{eqn:eq4}) – the convection – and accounts for mass transfer induced by the bulk fluid movement. In very high conductivity media, the convection magnitude may vastly exceed the DEP force, which complicates the depletion region visualization. The second factor that may obscure the depletion region in experiments is an inappropriate choice of the DEP electrode geometry. Indeed, the electrodes must ensure sufficient space between adjacent DEP traps to prevent the intersection of their depletion regions.

In summary, we have demonstrated that the experimental visualization of the equilibrium between particle diffusion and DEP translation can be utilized to investigate the DEP response of nanoscopic objects. As an example, we chose a colloidal solution of Au nanoparticles with different radii, calculated their trapping volumes from the concentration profiles measured in experiments, and compared the obtained results with numerical simulations. We also utilized these experimental concentration profiles to extract quantitative information on the system under test – here the Au nanoparticle radii. An excellent agreement was found between simulations and experiments, indicating the robustness of the proposed technique, which can be useful to investigate a broad diversity of analytes, for which more sophisticated DEP models are required. These findings shall stimulate experimental efforts to investigate the DEP response of more complex nanoscale particulates and assist the theoretical advancements in this field, rendering DEP a more quantitative and versatile tool for manipulations at the micro– and nanoscales.

\begin{acknowledgments}
The authors thank Christian Santschi and Sergejs Boroviks for fruitful discussions on the experimental and simulation data. This research did not receive any specific funding.
\end{acknowledgments}

\bibliographystyle{apsrev4-1}
\bibliography{references}

\begin{thebibliography}{57}%
\makeatletter
\providecommand \@ifxundefined [1]{%
 \@ifx{#1\undefined}
}%
\providecommand \@ifnum [1]{%
 \ifnum #1\expandafter \@firstoftwo
 \else \expandafter \@secondoftwo
 \fi
}%
\providecommand \@ifx [1]{%
 \ifx #1\expandafter \@firstoftwo
 \else \expandafter \@secondoftwo
 \fi
}%
\providecommand \natexlab [1]{#1}%
\providecommand \enquote  [1]{``#1''}%
\providecommand \bibnamefont  [1]{#1}%
\providecommand \bibfnamefont [1]{#1}%
\providecommand \citenamefont [1]{#1}%
\providecommand \href@noop [0]{\@secondoftwo}%
\providecommand \href [0]{\begingroup \@sanitize@url \@href}%
\providecommand \@href[1]{\@@startlink{#1}\@@href}%
\providecommand \@@href[1]{\endgroup#1\@@endlink}%
\providecommand \@sanitize@url [0]{\catcode `\\12\catcode `\$12\catcode `\&12\catcode `\#12\catcode `\^12\catcode `\_12\catcode `\%12\relax}%
\providecommand \@@startlink[1]{}%
\providecommand \@@endlink[0]{}%
\providecommand \url  [0]{\begingroup\@sanitize@url \@url }%
\providecommand \@url [1]{\endgroup\@href {#1}{\urlprefix }}%
\providecommand \urlprefix  [0]{URL }%
\providecommand \Eprint [0]{\href }%
\providecommand \doibase [0]{http://dx.doi.org/}%
\providecommand \selectlanguage [0]{\@gobble}%
\providecommand \bibinfo  [0]{\@secondoftwo}%
\providecommand \bibfield  [0]{\@secondoftwo}%
\providecommand \translation [1]{[#1]}%
\providecommand \BibitemOpen [0]{}%
\providecommand \bibitemStop [0]{}%
\providecommand \bibitemNoStop [0]{.\EOS\space}%
\providecommand \EOS [0]{\spacefactor3000\relax}%
\providecommand \BibitemShut  [1]{\csname bibitem#1\endcsname}%
\let\auto@bib@innerbib\@empty
\bibitem [{\citenamefont {Ramos}\ \emph {et~al.}(1998)\citenamefont {Ramos}, \citenamefont {Morgan}, \citenamefont {Green},\ and\ \citenamefont {Castellanos}}]{ramos_ac_1998}%
  \BibitemOpen
  \bibfield  {author} {\bibinfo {author} {\bibfnamefont {A.}~\bibnamefont {Ramos}}, \bibinfo {author} {\bibfnamefont {H.}~\bibnamefont {Morgan}}, \bibinfo {author} {\bibfnamefont {N.~G.}\ \bibnamefont {Green}}, \ and\ \bibinfo {author} {\bibfnamefont {A.}~\bibnamefont {Castellanos}},\ }\href {\doibase 10.1088/0022-3727/31/18/021} {\bibfield  {journal} {\bibinfo  {journal} {Journal of Physics D: Applied Physics}\ }\textbf {\bibinfo {volume} {31}},\ \bibinfo {pages} {2338} (\bibinfo {year} {1998})}\BibitemShut {NoStop}%
\bibitem [{\citenamefont {Schoch}\ \emph {et~al.}(2008)\citenamefont {Schoch}, \citenamefont {Han},\ and\ \citenamefont {Renaud}}]{schoch_transport_2008}%
  \BibitemOpen
  \bibfield  {author} {\bibinfo {author} {\bibfnamefont {R.~B.}\ \bibnamefont {Schoch}}, \bibinfo {author} {\bibfnamefont {J.}~\bibnamefont {Han}}, \ and\ \bibinfo {author} {\bibfnamefont {P.}~\bibnamefont {Renaud}},\ }\href {\doibase 10.1103/RevModPhys.80.839} {\bibfield  {journal} {\bibinfo  {journal} {Reviews of Modern Physics}\ }\textbf {\bibinfo {volume} {80}},\ \bibinfo {pages} {839} (\bibinfo {year} {2008})}\BibitemShut {NoStop}%
\bibitem [{\citenamefont {Squires}\ and\ \citenamefont {Quake}(2005)}]{squires_microfluidics_2005}%
  \BibitemOpen
  \bibfield  {author} {\bibinfo {author} {\bibfnamefont {T.~M.}\ \bibnamefont {Squires}}\ and\ \bibinfo {author} {\bibfnamefont {S.~R.}\ \bibnamefont {Quake}},\ }\href {\doibase 10.1103/RevModPhys.77.977} {\bibfield  {journal} {\bibinfo  {journal} {Reviews of Modern Physics}\ }\textbf {\bibinfo {volume} {77}},\ \bibinfo {pages} {977} (\bibinfo {year} {2005})}\BibitemShut {NoStop}%
\bibitem [{\citenamefont {Riccardi}\ and\ \citenamefont {Martin}(2023)}]{riccardi_electromagnetic_2023}%
  \BibitemOpen
  \bibfield  {author} {\bibinfo {author} {\bibfnamefont {M.}~\bibnamefont {Riccardi}}\ and\ \bibinfo {author} {\bibfnamefont {O.~J.~F.}\ \bibnamefont {Martin}},\ }\href {\doibase 10.1021/acs.chemrev.2c00576} {\bibfield  {journal} {\bibinfo  {journal} {Chemical Reviews}\ }\textbf {\bibinfo {volume} {123}},\ \bibinfo {pages} {1680} (\bibinfo {year} {2023})},\ \bibinfo {note} {publisher: American Chemical Society}\BibitemShut {NoStop}%
\bibitem [{\citenamefont {Yuan}\ \emph {et~al.}(2007)\citenamefont {Yuan}, \citenamefont {Garcia}, \citenamefont {Lopez},\ and\ \citenamefont {Petsev}}]{yuan_electrokinetic_2007}%
  \BibitemOpen
  \bibfield  {author} {\bibinfo {author} {\bibfnamefont {Z.}~\bibnamefont {Yuan}}, \bibinfo {author} {\bibfnamefont {A.~L.}\ \bibnamefont {Garcia}}, \bibinfo {author} {\bibfnamefont {G.~P.}\ \bibnamefont {Lopez}}, \ and\ \bibinfo {author} {\bibfnamefont {D.~N.}\ \bibnamefont {Petsev}},\ }\href {\doibase 10.1002/elps.200600612} {\bibfield  {journal} {\bibinfo  {journal} {Electrophoresis}\ }\textbf {\bibinfo {volume} {28}},\ \bibinfo {pages} {595} (\bibinfo {year} {2007})}\BibitemShut {NoStop}%
\bibitem [{\citenamefont {Llorente}\ \emph {et~al.}(2014)\citenamefont {Llorente}, \citenamefont {Fajardo},\ and\ \citenamefont {Bastidas}}]{llorente_applications_2014}%
  \BibitemOpen
  \bibfield  {author} {\bibinfo {author} {\bibfnamefont {I.}~\bibnamefont {Llorente}}, \bibinfo {author} {\bibfnamefont {S.}~\bibnamefont {Fajardo}}, \ and\ \bibinfo {author} {\bibfnamefont {J.~M.}\ \bibnamefont {Bastidas}},\ }\href {\doibase 10.1007/s10008-013-2267-0} {\bibfield  {journal} {\bibinfo  {journal} {Journal of Solid State Electrochemistry}\ }\textbf {\bibinfo {volume} {18}},\ \bibinfo {pages} {293} (\bibinfo {year} {2014})}\BibitemShut {NoStop}%
\bibitem [{\citenamefont {Mir}\ \emph {et~al.}(2011)\citenamefont {Mir}, \citenamefont {Martínez-Rodríguez}, \citenamefont {Castillo-Fernández}, \citenamefont {Homs-Corbera},\ and\ \citenamefont {Samitier}}]{mir_electrokinetic_2011}%
  \BibitemOpen
  \bibfield  {author} {\bibinfo {author} {\bibfnamefont {M.}~\bibnamefont {Mir}}, \bibinfo {author} {\bibfnamefont {S.}~\bibnamefont {Martínez-Rodríguez}}, \bibinfo {author} {\bibfnamefont {O.}~\bibnamefont {Castillo-Fernández}}, \bibinfo {author} {\bibfnamefont {A.}~\bibnamefont {Homs-Corbera}}, \ and\ \bibinfo {author} {\bibfnamefont {J.}~\bibnamefont {Samitier}},\ }\href {\doibase 10.1002/elps.201000487} {\bibfield  {journal} {\bibinfo  {journal} {Electrophoresis}\ }\textbf {\bibinfo {volume} {32}},\ \bibinfo {pages} {811} (\bibinfo {year} {2011})}\BibitemShut {NoStop}%
\bibitem [{\citenamefont {Pethig}(2010)}]{pethig_dielectrophoresis_2010}%
  \BibitemOpen
  \bibfield  {author} {\bibinfo {author} {\bibfnamefont {R.}~\bibnamefont {Pethig}},\ }\href {\doibase 10.1063/1.3456626} {\bibfield  {journal} {\bibinfo  {journal} {Biomicrofluidics}\ }\textbf {\bibinfo {volume} {4}},\ \bibinfo {pages} {022811} (\bibinfo {year} {2010})},\ \bibinfo {note} {publisher: American Institute of PhysicsAIP}\BibitemShut {NoStop}%
\bibitem [{\citenamefont {Pethig}(2017)}]{pethig_dielectrophoresis_2017}%
  \BibitemOpen
  \bibfield  {author} {\bibinfo {author} {\bibfnamefont {R.}~\bibnamefont {Pethig}},\ }\href {\doibase 10.22456/2527-2616.75900} {\emph {\bibinfo {title} {Dielectrophoresis: theory, methodology and biological applications}}}\ (\bibinfo  {publisher} {John Wiley \& Sons},\ \bibinfo {address} {Hoboken, NJ},\ \bibinfo {year} {2017})\ \bibinfo {note} {iSSN: 00136654}\BibitemShut {NoStop}%
\bibitem [{\citenamefont {Camacho-Alanis}\ \emph {et~al.}(2012)\citenamefont {Camacho-Alanis}, \citenamefont {Gan},\ and\ \citenamefont {Ros}}]{camacho-alanis_transitioning_2012}%
  \BibitemOpen
  \bibfield  {author} {\bibinfo {author} {\bibfnamefont {F.}~\bibnamefont {Camacho-Alanis}}, \bibinfo {author} {\bibfnamefont {L.}~\bibnamefont {Gan}}, \ and\ \bibinfo {author} {\bibfnamefont {A.}~\bibnamefont {Ros}},\ }\href {\doibase 10.1016/j.snb.2012.07.080} {\bibfield  {journal} {\bibinfo  {journal} {Sensors and Actuators, B: Chemical}\ }\textbf {\bibinfo {volume} {173}},\ \bibinfo {pages} {668} (\bibinfo {year} {2012})},\ \bibinfo {note} {publisher: Elsevier B.V.}\BibitemShut {Stop}%
\bibitem [{\citenamefont {Hughes}\ and\ \citenamefont {Morgan}(1998)}]{hughes_dielectrophoretic_1998}%
  \BibitemOpen
  \bibfield  {author} {\bibinfo {author} {\bibfnamefont {M.~P.}\ \bibnamefont {Hughes}}\ and\ \bibinfo {author} {\bibfnamefont {H.}~\bibnamefont {Morgan}},\ }\href {\doibase 10.1088/0022-3727/31/17/020} {\bibfield  {journal} {\bibinfo  {journal} {Journal of Physics D: Applied Physics}\ }\textbf {\bibinfo {volume} {31}},\ \bibinfo {pages} {2205} (\bibinfo {year} {1998})}\BibitemShut {NoStop}%
\bibitem [{\citenamefont {Hölzel}\ \emph {et~al.}(2005)\citenamefont {Hölzel}, \citenamefont {Calander}, \citenamefont {Chiragwandi}, \citenamefont {Willander},\ and\ \citenamefont {Bier}}]{holzel_trapping_2005}%
  \BibitemOpen
  \bibfield  {author} {\bibinfo {author} {\bibfnamefont {R.}~\bibnamefont {Hölzel}}, \bibinfo {author} {\bibfnamefont {N.}~\bibnamefont {Calander}}, \bibinfo {author} {\bibfnamefont {Z.}~\bibnamefont {Chiragwandi}}, \bibinfo {author} {\bibfnamefont {M.}~\bibnamefont {Willander}}, \ and\ \bibinfo {author} {\bibfnamefont {F.~F.}\ \bibnamefont {Bier}},\ }\href {\doibase 10.1103/PhysRevLett.95.128102} {\bibfield  {journal} {\bibinfo  {journal} {Physical Review Letters}\ }\textbf {\bibinfo {volume} {95}},\ \bibinfo {pages} {18} (\bibinfo {year} {2005})}\BibitemShut {NoStop}%
\bibitem [{\citenamefont {Liu}\ and\ \citenamefont {Hayes}(2021)}]{liu_orders--magnitude_2021}%
  \BibitemOpen
  \bibfield  {author} {\bibinfo {author} {\bibfnamefont {Y.}~\bibnamefont {Liu}}\ and\ \bibinfo {author} {\bibfnamefont {M.~A.}\ \bibnamefont {Hayes}},\ }\href {\doibase 10.1021/acs.analchem.0c02763} {\bibfield  {journal} {\bibinfo  {journal} {Analytical Chemistry}\ } (\bibinfo {year} {2021}),\ 10.1021/acs.analchem.0c02763}\BibitemShut {NoStop}%
\bibitem [{\citenamefont {Regtmeier}\ \emph {et~al.}(2007)\citenamefont {Regtmeier}, \citenamefont {Duong}, \citenamefont {Eichhorn}, \citenamefont {Anselmetti},\ and\ \citenamefont {Ros}}]{regtmeier_dielectrophoretic_2007}%
  \BibitemOpen
  \bibfield  {author} {\bibinfo {author} {\bibfnamefont {J.}~\bibnamefont {Regtmeier}}, \bibinfo {author} {\bibfnamefont {T.~T.}\ \bibnamefont {Duong}}, \bibinfo {author} {\bibfnamefont {R.}~\bibnamefont {Eichhorn}}, \bibinfo {author} {\bibfnamefont {D.}~\bibnamefont {Anselmetti}}, \ and\ \bibinfo {author} {\bibfnamefont {A.}~\bibnamefont {Ros}},\ }\href {\doibase 10.1021/ac062431r} {\bibfield  {journal} {\bibinfo  {journal} {Analytical Chemistry}\ }\textbf {\bibinfo {volume} {79}},\ \bibinfo {pages} {3925} (\bibinfo {year} {2007})},\ \bibinfo {note} {publisher: American Chemical Society}\BibitemShut {NoStop}%
\bibitem [{\citenamefont {Yunus}\ \emph {et~al.}(2013)\citenamefont {Yunus}, \citenamefont {Nili},\ and\ \citenamefont {Green}}]{yunus_continuous_2013}%
  \BibitemOpen
  \bibfield  {author} {\bibinfo {author} {\bibfnamefont {N.~A.~M.}\ \bibnamefont {Yunus}}, \bibinfo {author} {\bibfnamefont {H.}~\bibnamefont {Nili}}, \ and\ \bibinfo {author} {\bibfnamefont {N.~G.}\ \bibnamefont {Green}},\ }\href {\doibase 10.1002/ELPS.201200466} {\bibfield  {journal} {\bibinfo  {journal} {Electrophoresis}\ }\textbf {\bibinfo {volume} {34}},\ \bibinfo {pages} {969} (\bibinfo {year} {2013})}\BibitemShut {NoStop}%
\bibitem [{\citenamefont {Liao}\ and\ \citenamefont {Chou}(2012)}]{liao_nanoscale_2012}%
  \BibitemOpen
  \bibfield  {author} {\bibinfo {author} {\bibfnamefont {K.~T.}\ \bibnamefont {Liao}}\ and\ \bibinfo {author} {\bibfnamefont {C.~F.}\ \bibnamefont {Chou}},\ }\href {\doibase 10.1021/ja3016523} {\bibfield  {journal} {\bibinfo  {journal} {Journal of the American Chemical Society}\ }\textbf {\bibinfo {volume} {134}},\ \bibinfo {pages} {8742} (\bibinfo {year} {2012})}\BibitemShut {NoStop}%
\bibitem [{\citenamefont {LaLonde}\ \emph {et~al.}(2015)\citenamefont {LaLonde}, \citenamefont {Romero-Creel}, \citenamefont {Saucedo-Espinosa},\ and\ \citenamefont {Lapizco-Encinas}}]{lalonde_isolation_2015}%
  \BibitemOpen
  \bibfield  {author} {\bibinfo {author} {\bibfnamefont {A.}~\bibnamefont {LaLonde}}, \bibinfo {author} {\bibfnamefont {M.~F.}\ \bibnamefont {Romero-Creel}}, \bibinfo {author} {\bibfnamefont {M.~A.}\ \bibnamefont {Saucedo-Espinosa}}, \ and\ \bibinfo {author} {\bibfnamefont {B.~H.}\ \bibnamefont {Lapizco-Encinas}},\ }\href {\doibase 10.1063/1.4936371} {\bibfield  {journal} {\bibinfo  {journal} {Biomicrofluidics}\ }\textbf {\bibinfo {volume} {9}},\ \bibinfo {pages} {064113} (\bibinfo {year} {2015})}\BibitemShut {NoStop}%
\bibitem [{\citenamefont {D'Amico}\ \emph {et~al.}(2017)\citenamefont {D'Amico}, \citenamefont {Ajami}, \citenamefont {Adachi}, \citenamefont {Gascoyne},\ and\ \citenamefont {Petrosino}}]{damico_isolation_2017}%
  \BibitemOpen
  \bibfield  {author} {\bibinfo {author} {\bibfnamefont {L.}~\bibnamefont {D'Amico}}, \bibinfo {author} {\bibfnamefont {N.~J.}\ \bibnamefont {Ajami}}, \bibinfo {author} {\bibfnamefont {J.~A.}\ \bibnamefont {Adachi}}, \bibinfo {author} {\bibfnamefont {P.~R.}\ \bibnamefont {Gascoyne}}, \ and\ \bibinfo {author} {\bibfnamefont {J.~F.}\ \bibnamefont {Petrosino}},\ }\href {\doibase 10.1039/C6LC01277A} {\bibfield  {journal} {\bibinfo  {journal} {Lab on a Chip}\ }\textbf {\bibinfo {volume} {17}},\ \bibinfo {pages} {1340} (\bibinfo {year} {2017})}\BibitemShut {NoStop}%
\bibitem [{\citenamefont {Nguyen}\ and\ \citenamefont {Jen}(2018)}]{nguyen_impedance_2018}%
  \BibitemOpen
  \bibfield  {author} {\bibinfo {author} {\bibfnamefont {N.-V.}\ \bibnamefont {Nguyen}}\ and\ \bibinfo {author} {\bibfnamefont {C.-P.}\ \bibnamefont {Jen}},\ }\href {\doibase 10.1016/j.bios.2018.08.059} {\bibfield  {journal} {\bibinfo  {journal} {Biosensors and Bioelectronics}\ }\textbf {\bibinfo {volume} {121}},\ \bibinfo {pages} {10} (\bibinfo {year} {2018})}\BibitemShut {NoStop}%
\bibitem [{\citenamefont {Pethig}(2019)}]{pethig_limitations_2019}%
  \BibitemOpen
  \bibfield  {author} {\bibinfo {author} {\bibfnamefont {R.}~\bibnamefont {Pethig}},\ }\href {\doibase 10.1002/elps.201900057} {\bibfield  {journal} {\bibinfo  {journal} {Electrophoresis}\ }\textbf {\bibinfo {volume} {40}},\ \bibinfo {pages} {2575} (\bibinfo {year} {2019})}\BibitemShut {NoStop}%
\bibitem [{\citenamefont {Pethig}(2022)}]{pethig_protein_2022}%
  \BibitemOpen
  \bibfield  {author} {\bibinfo {author} {\bibfnamefont {R.}~\bibnamefont {Pethig}},\ }\href {\doibase 10.3390/mi13020261} {\bibfield  {journal} {\bibinfo  {journal} {Micromachines}\ }\textbf {\bibinfo {volume} {13}},\ \bibinfo {pages} {261} (\bibinfo {year} {2022})}\BibitemShut {NoStop}%
\bibitem [{\citenamefont {Hölzel}\ and\ \citenamefont {Pethig}(2020)}]{holzel_protein_2020}%
  \BibitemOpen
  \bibfield  {author} {\bibinfo {author} {\bibfnamefont {R.}~\bibnamefont {Hölzel}}\ and\ \bibinfo {author} {\bibfnamefont {R.}~\bibnamefont {Pethig}},\ }\href {\doibase 10.3390/mi11050533} {\bibfield  {journal} {\bibinfo  {journal} {Micromachines}\ }\textbf {\bibinfo {volume} {11}},\ \bibinfo {pages} {533} (\bibinfo {year} {2020})}\BibitemShut {NoStop}%
\bibitem [{\citenamefont {Hölzel}\ and\ \citenamefont {Pethig}(2021)}]{holzel_protein_2021}%
  \BibitemOpen
  \bibfield  {author} {\bibinfo {author} {\bibfnamefont {R.}~\bibnamefont {Hölzel}}\ and\ \bibinfo {author} {\bibfnamefont {R.}~\bibnamefont {Pethig}},\ }\href {\doibase 10.1002/elps.202000255} {\bibfield  {journal} {\bibinfo  {journal} {Electrophoresis}\ }\textbf {\bibinfo {volume} {42}},\ \bibinfo {pages} {513} (\bibinfo {year} {2021})}\BibitemShut {NoStop}%
\bibitem [{\citenamefont {Seyedi}\ and\ \citenamefont {Matyushov}(2018)}]{seyedi_protein_2018}%
  \BibitemOpen
  \bibfield  {author} {\bibinfo {author} {\bibfnamefont {S.~S.}\ \bibnamefont {Seyedi}}\ and\ \bibinfo {author} {\bibfnamefont {D.~V.}\ \bibnamefont {Matyushov}},\ }\href {\doibase 10.1021/acs.jpcb.8b06864} {\bibfield  {journal} {\bibinfo  {journal} {Journal of Physical Chemistry B}\ }\textbf {\bibinfo {volume} {122}},\ \bibinfo {pages} {9119} (\bibinfo {year} {2018})}\BibitemShut {NoStop}%
\bibitem [{\citenamefont {Heyden}\ and\ \citenamefont {Matyushov}(2020)}]{heyden_dielectrophoresis_2020}%
  \BibitemOpen
  \bibfield  {author} {\bibinfo {author} {\bibfnamefont {M.}~\bibnamefont {Heyden}}\ and\ \bibinfo {author} {\bibfnamefont {D.~V.}\ \bibnamefont {Matyushov}},\ }\href {\doibase 10.1021/acs.jpcb.0c09007} {\bibfield  {journal} {\bibinfo  {journal} {Journal of Physical Chemistry B}\ }\textbf {\bibinfo {volume} {124}},\ \bibinfo {pages} {11634} (\bibinfo {year} {2020})}\BibitemShut {NoStop}%
\bibitem [{\citenamefont {Hoettges}(2010)}]{hoettges_dielectrophoresis_2010}%
  \BibitemOpen
  \bibfield  {author} {\bibinfo {author} {\bibfnamefont {K.~F.}\ \bibnamefont {Hoettges}},\ }in\ \href {\doibase 10.1007/978-1-60327-106-6_8} {\emph {\bibinfo {booktitle} {Microengineering in {Biotechnology}}}},\ \bibinfo {series and number} {Methods in {Molecular} {Biology}},\ \bibinfo {editor} {edited by\ \bibinfo {editor} {\bibfnamefont {M.~P.}\ \bibnamefont {Hughes}}\ and\ \bibinfo {editor} {\bibfnamefont {K.~F.}\ \bibnamefont {Hoettges}}}\ (\bibinfo  {publisher} {Humana Press},\ \bibinfo {address} {Totowa, NJ},\ \bibinfo {year} {2010})\ pp.\ \bibinfo {pages} {183--198}\BibitemShut {NoStop}%
\bibitem [{\citenamefont {Watarai}\ \emph {et~al.}(1997)\citenamefont {Watarai}, \citenamefont {Sakamoto},\ and\ \citenamefont {Tsukahara}}]{watarai_situ_1997}%
  \BibitemOpen
  \bibfield  {author} {\bibinfo {author} {\bibfnamefont {H.}~\bibnamefont {Watarai}}, \bibinfo {author} {\bibfnamefont {T.}~\bibnamefont {Sakamoto}}, \ and\ \bibinfo {author} {\bibfnamefont {S.}~\bibnamefont {Tsukahara}},\ }\href {\doibase 10.1021/la961057v} {\bibfield  {journal} {\bibinfo  {journal} {Langmuir}\ }\textbf {\bibinfo {volume} {13}},\ \bibinfo {pages} {2417} (\bibinfo {year} {1997})}\BibitemShut {NoStop}%
\bibitem [{\citenamefont {Huang}\ \emph {et~al.}(1992)\citenamefont {Huang}, \citenamefont {Holzel}, \citenamefont {Pethig},\ and\ \citenamefont {Wang}}]{huang_differences_1992}%
  \BibitemOpen
  \bibfield  {author} {\bibinfo {author} {\bibfnamefont {Y.}~\bibnamefont {Huang}}, \bibinfo {author} {\bibfnamefont {R.}~\bibnamefont {Holzel}}, \bibinfo {author} {\bibfnamefont {R.}~\bibnamefont {Pethig}}, \ and\ \bibinfo {author} {\bibfnamefont {X.-B.}\ \bibnamefont {Wang}},\ }\href {\doibase 10.1088/0031-9155/37/7/003} {\bibfield  {journal} {\bibinfo  {journal} {Physics in Medicine \& Biology}\ }\textbf {\bibinfo {volume} {37}},\ \bibinfo {pages} {1499} (\bibinfo {year} {1992})}\BibitemShut {NoStop}%
\bibitem [{\citenamefont {Kralj}\ \emph {et~al.}(2006)\citenamefont {Kralj}, \citenamefont {Lis}, \citenamefont {Schmidt},\ and\ \citenamefont {Jensen}}]{kralj_continuous_2006}%
  \BibitemOpen
  \bibfield  {author} {\bibinfo {author} {\bibfnamefont {J.~G.}\ \bibnamefont {Kralj}}, \bibinfo {author} {\bibfnamefont {M.~T.~W.}\ \bibnamefont {Lis}}, \bibinfo {author} {\bibfnamefont {M.~A.}\ \bibnamefont {Schmidt}}, \ and\ \bibinfo {author} {\bibfnamefont {K.~F.}\ \bibnamefont {Jensen}},\ }\href {\doibase 10.1021/ac0601314} {\bibfield  {journal} {\bibinfo  {journal} {Analytical Chemistry}\ }\textbf {\bibinfo {volume} {78}},\ \bibinfo {pages} {5019} (\bibinfo {year} {2006})}\BibitemShut {NoStop}%
\bibitem [{\citenamefont {Ai}\ \emph {et~al.}(2009)\citenamefont {Ai}, \citenamefont {Joo}, \citenamefont {Jiang}, \citenamefont {Xuan},\ and\ \citenamefont {Qian}}]{ai_transient_2009}%
  \BibitemOpen
  \bibfield  {author} {\bibinfo {author} {\bibfnamefont {Y.}~\bibnamefont {Ai}}, \bibinfo {author} {\bibfnamefont {S.~W.}\ \bibnamefont {Joo}}, \bibinfo {author} {\bibfnamefont {Y.}~\bibnamefont {Jiang}}, \bibinfo {author} {\bibfnamefont {X.}~\bibnamefont {Xuan}}, \ and\ \bibinfo {author} {\bibfnamefont {S.}~\bibnamefont {Qian}},\ }\href {\doibase 10.1002/elps.200800792} {\bibfield  {journal} {\bibinfo  {journal} {Electrophoresis}\ }\textbf {\bibinfo {volume} {30}},\ \bibinfo {pages} {2499} (\bibinfo {year} {2009})}\BibitemShut {NoStop}%
\bibitem [{\citenamefont {Lemons}\ and\ \citenamefont {Gythiel}(1997)}]{lemons_paul_1997}%
  \BibitemOpen
  \bibfield  {author} {\bibinfo {author} {\bibfnamefont {D.~S.}\ \bibnamefont {Lemons}}\ and\ \bibinfo {author} {\bibfnamefont {A.}~\bibnamefont {Gythiel}},\ }\href {\doibase 10.1119/1.18725} {\bibfield  {journal} {\bibinfo  {journal} {American Journal of Physics}\ }\textbf {\bibinfo {volume} {65}},\ \bibinfo {pages} {1079} (\bibinfo {year} {1997})}\BibitemShut {NoStop}%
\bibitem [{\citenamefont {Langevin}(1908)}]{langevin_sur_1908}%
  \BibitemOpen
  \bibfield  {author} {\bibinfo {author} {\bibfnamefont {P.}~\bibnamefont {Langevin}},\ }\href@noop {} {\bibfield  {journal} {\bibinfo  {journal} {C. R. Acad. Sci. Paris}\ }\textbf {\bibinfo {volume} {146}},\ \bibinfo {pages} {530} (\bibinfo {year} {1908})}\BibitemShut {NoStop}%
\bibitem [{\citenamefont {Park}\ \emph {et~al.}(2014)\citenamefont {Park}, \citenamefont {Park}, \citenamefont {Yoon}, \citenamefont {Lee},\ and\ \citenamefont {Kim}}]{park_direct_2014}%
  \BibitemOpen
  \bibfield  {author} {\bibinfo {author} {\bibfnamefont {I.~S.}\ \bibnamefont {Park}}, \bibinfo {author} {\bibfnamefont {S.~H.}\ \bibnamefont {Park}}, \bibinfo {author} {\bibfnamefont {D.~S.}\ \bibnamefont {Yoon}}, \bibinfo {author} {\bibfnamefont {S.~W.}\ \bibnamefont {Lee}}, \ and\ \bibinfo {author} {\bibfnamefont {B.-M.}\ \bibnamefont {Kim}},\ }\href {\doibase 10.1063/1.4895115} {\bibfield  {journal} {\bibinfo  {journal} {Applied Physics Letters}\ }\textbf {\bibinfo {volume} {105}},\ \bibinfo {pages} {103701} (\bibinfo {year} {2014})}\BibitemShut {NoStop}%
\bibitem [{\citenamefont {Hong}\ \emph {et~al.}(2010)\citenamefont {Hong}, \citenamefont {Pyo}, \citenamefont {Baek}, \citenamefont {Lee}, \citenamefont {Yoon}, \citenamefont {No},\ and\ \citenamefont {Kim}}]{hong_quantitative_2010}%
  \BibitemOpen
  \bibfield  {author} {\bibinfo {author} {\bibfnamefont {Y.}~\bibnamefont {Hong}}, \bibinfo {author} {\bibfnamefont {J.-W.}\ \bibnamefont {Pyo}}, \bibinfo {author} {\bibfnamefont {S.~H.}\ \bibnamefont {Baek}}, \bibinfo {author} {\bibfnamefont {S.~W.}\ \bibnamefont {Lee}}, \bibinfo {author} {\bibfnamefont {D.~S.}\ \bibnamefont {Yoon}}, \bibinfo {author} {\bibfnamefont {K.}~\bibnamefont {No}}, \ and\ \bibinfo {author} {\bibfnamefont {B.-M.}\ \bibnamefont {Kim}},\ }\href {\doibase 10.1364/OL.35.002493} {\bibfield  {journal} {\bibinfo  {journal} {Optics Letters}\ }\textbf {\bibinfo {volume} {35}},\ \bibinfo {pages} {2493} (\bibinfo {year} {2010})}\BibitemShut {NoStop}%
\bibitem [{\citenamefont {Jeon}\ \emph {et~al.}(2017)\citenamefont {Jeon}, \citenamefont {Lee}, \citenamefont {Yoon},\ and\ \citenamefont {Kim}}]{jeon_dielectrophoretic_2017}%
  \BibitemOpen
  \bibfield  {author} {\bibinfo {author} {\bibfnamefont {H.-J.}\ \bibnamefont {Jeon}}, \bibinfo {author} {\bibfnamefont {H.}~\bibnamefont {Lee}}, \bibinfo {author} {\bibfnamefont {D.~S.}\ \bibnamefont {Yoon}}, \ and\ \bibinfo {author} {\bibfnamefont {B.-M.}\ \bibnamefont {Kim}},\ }\href {\doibase 10.1007/s13534-017-0041-4} {\bibfield  {journal} {\bibinfo  {journal} {Biomedical Engineering Letters}\ }\textbf {\bibinfo {volume} {7}},\ \bibinfo {pages} {317} (\bibinfo {year} {2017})}\BibitemShut {NoStop}%
\bibitem [{\citenamefont {Imasato}\ and\ \citenamefont {Yamakawa}(2008)}]{imasato_measurement_2008}%
  \BibitemOpen
  \bibfield  {author} {\bibinfo {author} {\bibfnamefont {H.}~\bibnamefont {Imasato}}\ and\ \bibinfo {author} {\bibfnamefont {T.}~\bibnamefont {Yamakawa}},\ }\href {\doibase 10.2198/jelectroph.52.1} {\bibfield  {journal} {\bibinfo  {journal} {Journal of Electrophoresis}\ }\textbf {\bibinfo {volume} {52}},\ \bibinfo {pages} {1} (\bibinfo {year} {2008})}\BibitemShut {NoStop}%
\bibitem [{\citenamefont {Lu}\ \emph {et~al.}(2020)\citenamefont {Lu}, \citenamefont {Sun}, \citenamefont {Kao}, \citenamefont {Hung},\ and\ \citenamefont {Juang}}]{lu_dielectrophoretic_2020}%
  \BibitemOpen
  \bibfield  {author} {\bibinfo {author} {\bibfnamefont {Y.-W.}\ \bibnamefont {Lu}}, \bibinfo {author} {\bibfnamefont {C.}~\bibnamefont {Sun}}, \bibinfo {author} {\bibfnamefont {Y.-C.}\ \bibnamefont {Kao}}, \bibinfo {author} {\bibfnamefont {C.-L.}\ \bibnamefont {Hung}}, \ and\ \bibinfo {author} {\bibfnamefont {J.-Y.}\ \bibnamefont {Juang}},\ }\href {\doibase 10.3390/nano10071364} {\bibfield  {journal} {\bibinfo  {journal} {Nanomaterials}\ }\textbf {\bibinfo {volume} {10}},\ \bibinfo {pages} {1364} (\bibinfo {year} {2020})}\BibitemShut {NoStop}%
\bibitem [{\citenamefont {Su}\ \emph {et~al.}(2013)\citenamefont {Su}, \citenamefont {L. Prieto},\ and\ \citenamefont {Voldman}}]{su_rapid_2013}%
  \BibitemOpen
  \bibfield  {author} {\bibinfo {author} {\bibfnamefont {H.-W.}\ \bibnamefont {Su}}, \bibinfo {author} {\bibfnamefont {J.}~\bibnamefont {L. Prieto}}, \ and\ \bibinfo {author} {\bibfnamefont {J.}~\bibnamefont {Voldman}},\ }\href {\doibase 10.1039/C3LC50392E} {\bibfield  {journal} {\bibinfo  {journal} {Lab on a Chip}\ }\textbf {\bibinfo {volume} {13}},\ \bibinfo {pages} {4109} (\bibinfo {year} {2013})}\BibitemShut {NoStop}%
\bibitem [{\citenamefont {Zavatski}\ \emph {et~al.}(2023)\citenamefont {Zavatski}, \citenamefont {Bandarenka},\ and\ \citenamefont {Martin}}]{zavatski_protein_2023}%
  \BibitemOpen
  \bibfield  {author} {\bibinfo {author} {\bibfnamefont {S.}~\bibnamefont {Zavatski}}, \bibinfo {author} {\bibfnamefont {H.}~\bibnamefont {Bandarenka}}, \ and\ \bibinfo {author} {\bibfnamefont {O.~J.~F.}\ \bibnamefont {Martin}},\ }\href {\doibase 10.1021/acs.analchem.2c04708} {\bibfield  {journal} {\bibinfo  {journal} {Analytical Chemistry}\ }\textbf {\bibinfo {volume} {95}},\ \bibinfo {pages} {2958} (\bibinfo {year} {2023})}\BibitemShut {NoStop}%
\bibitem [{\citenamefont {Labeed}\ \emph {et~al.}(2003)\citenamefont {Labeed}, \citenamefont {Coley}, \citenamefont {Thomas},\ and\ \citenamefont {Hughes}}]{labeed_assessment_2003}%
  \BibitemOpen
  \bibfield  {author} {\bibinfo {author} {\bibfnamefont {F.~H.}\ \bibnamefont {Labeed}}, \bibinfo {author} {\bibfnamefont {H.~M.}\ \bibnamefont {Coley}}, \bibinfo {author} {\bibfnamefont {H.}~\bibnamefont {Thomas}}, \ and\ \bibinfo {author} {\bibfnamefont {M.~P.}\ \bibnamefont {Hughes}},\ }\href {\doibase 10.1016/S0006-3495(03)74630-X} {\bibfield  {journal} {\bibinfo  {journal} {Biophysical Journal}\ }\textbf {\bibinfo {volume} {85}},\ \bibinfo {pages} {2028} (\bibinfo {year} {2003})}\BibitemShut {NoStop}%
\bibitem [{\citenamefont {Labeed}\ \emph {et~al.}(2006)\citenamefont {Labeed}, \citenamefont {Coley},\ and\ \citenamefont {Hughes}}]{labeed_differences_2006}%
  \BibitemOpen
  \bibfield  {author} {\bibinfo {author} {\bibfnamefont {F.~H.}\ \bibnamefont {Labeed}}, \bibinfo {author} {\bibfnamefont {H.~M.}\ \bibnamefont {Coley}}, \ and\ \bibinfo {author} {\bibfnamefont {M.~P.}\ \bibnamefont {Hughes}},\ }\href {\doibase 10.1016/j.bbagen.2006.01.018} {\bibfield  {journal} {\bibinfo  {journal} {Biochimica et Biophysica Acta (BBA) - General Subjects}\ }\textbf {\bibinfo {volume} {1760}},\ \bibinfo {pages} {922} (\bibinfo {year} {2006})}\BibitemShut {NoStop}%
\bibitem [{\citenamefont {Markx}\ \emph {et~al.}(1994)\citenamefont {Markx}, \citenamefont {Huang}, \citenamefont {Zhou},\ and\ \citenamefont {Pethig}}]{markx_dielectrophoretic_1994}%
  \BibitemOpen
  \bibfield  {author} {\bibinfo {author} {\bibfnamefont {G.~H.}\ \bibnamefont {Markx}}, \bibinfo {author} {\bibfnamefont {Y.}~\bibnamefont {Huang}}, \bibinfo {author} {\bibfnamefont {X.-F.}\ \bibnamefont {Zhou}}, \ and\ \bibinfo {author} {\bibfnamefont {R.}~\bibnamefont {Pethig}},\ }\href {\doibase 10.1099/00221287-140-3-585} {\bibfield  {journal} {\bibinfo  {journal} {Microbiology}\ }\textbf {\bibinfo {volume} {140}},\ \bibinfo {pages} {585} (\bibinfo {year} {1994})}\BibitemShut {NoStop}%
\bibitem [{\citenamefont {Hübner}\ \emph {et~al.}(2003)\citenamefont {Hübner}, \citenamefont {Hoettges},\ and\ \citenamefont {Hughes}}]{hubner_water_2003}%
  \BibitemOpen
  \bibfield  {author} {\bibinfo {author} {\bibfnamefont {Y.}~\bibnamefont {Hübner}}, \bibinfo {author} {\bibfnamefont {K.~F.}\ \bibnamefont {Hoettges}}, \ and\ \bibinfo {author} {\bibfnamefont {M.~P.}\ \bibnamefont {Hughes}},\ }\href {\doibase 10.1039/B309131G} {\bibfield  {journal} {\bibinfo  {journal} {Journal of Environmental Monitoring}\ }\textbf {\bibinfo {volume} {5}},\ \bibinfo {pages} {861} (\bibinfo {year} {2003})}\BibitemShut {NoStop}%
\bibitem [{\citenamefont {Hughes}\ \emph {et~al.}(1998)\citenamefont {Hughes}, \citenamefont {Morgan}, \citenamefont {Rixon}, \citenamefont {Burt},\ and\ \citenamefont {Pethig}}]{hughes_manipulation_1998}%
  \BibitemOpen
  \bibfield  {author} {\bibinfo {author} {\bibfnamefont {M.~P.}\ \bibnamefont {Hughes}}, \bibinfo {author} {\bibfnamefont {H.}~\bibnamefont {Morgan}}, \bibinfo {author} {\bibfnamefont {F.~J.}\ \bibnamefont {Rixon}}, \bibinfo {author} {\bibfnamefont {J.~P.~H.}\ \bibnamefont {Burt}}, \ and\ \bibinfo {author} {\bibfnamefont {R.}~\bibnamefont {Pethig}},\ }\href {\doibase 10.1016/S0304-4165(98)00058-0} {\bibfield  {journal} {\bibinfo  {journal} {Biochimica et Biophysica Acta (BBA) - General Subjects}\ }\textbf {\bibinfo {volume} {1425}},\ \bibinfo {pages} {119} (\bibinfo {year} {1998})}\BibitemShut {NoStop}%
\bibitem [{\citenamefont {Green}\ \emph {et~al.}(1997)\citenamefont {Green}, \citenamefont {Morgan},\ and\ \citenamefont {Milner}}]{green_manipulation_1997}%
  \BibitemOpen
  \bibfield  {author} {\bibinfo {author} {\bibfnamefont {N.~G.}\ \bibnamefont {Green}}, \bibinfo {author} {\bibfnamefont {H.}~\bibnamefont {Morgan}}, \ and\ \bibinfo {author} {\bibfnamefont {J.~J.}\ \bibnamefont {Milner}},\ }\href {\doibase 10.1016/S0165-022X(97)00033-X} {\bibfield  {journal} {\bibinfo  {journal} {Journal of Biochemical and Biophysical Methods}\ }\textbf {\bibinfo {volume} {35}},\ \bibinfo {pages} {89} (\bibinfo {year} {1997})}\BibitemShut {NoStop}%
\bibitem [{\citenamefont {Schnelle}\ \emph {et~al.}(1996)\citenamefont {Schnelle}, \citenamefont {Müller}, \citenamefont {Fiedler}, \citenamefont {Shirley}, \citenamefont {Ludwig}, \citenamefont {Herrmann}, \citenamefont {Fuhr}, \citenamefont {Wagner},\ and\ \citenamefont {Zimmermann}}]{schnelle_trapping_1996}%
  \BibitemOpen
  \bibfield  {author} {\bibinfo {author} {\bibfnamefont {T.}~\bibnamefont {Schnelle}}, \bibinfo {author} {\bibfnamefont {T.}~\bibnamefont {Müller}}, \bibinfo {author} {\bibfnamefont {S.}~\bibnamefont {Fiedler}}, \bibinfo {author} {\bibfnamefont {S.~G.}\ \bibnamefont {Shirley}}, \bibinfo {author} {\bibfnamefont {K.}~\bibnamefont {Ludwig}}, \bibinfo {author} {\bibfnamefont {A.}~\bibnamefont {Herrmann}}, \bibinfo {author} {\bibfnamefont {G.}~\bibnamefont {Fuhr}}, \bibinfo {author} {\bibfnamefont {B.}~\bibnamefont {Wagner}}, \ and\ \bibinfo {author} {\bibfnamefont {U.}~\bibnamefont {Zimmermann}},\ }\href {\doibase 10.1007/BF01143058} {\bibfield  {journal} {\bibinfo  {journal} {Naturwissenschaften}\ }\textbf {\bibinfo {volume} {83}},\ \bibinfo {pages} {172} (\bibinfo {year} {1996})}\BibitemShut {NoStop}%
\bibitem [{\citenamefont {Jones}\ and\ \citenamefont {Bliss}(2008)}]{jones_bubble_2008}%
  \BibitemOpen
  \bibfield  {author} {\bibinfo {author} {\bibfnamefont {T.~B.}\ \bibnamefont {Jones}}\ and\ \bibinfo {author} {\bibfnamefont {G.~W.}\ \bibnamefont {Bliss}},\ }\href {\doibase 10.1063/1.323806} {\bibfield  {journal} {\bibinfo  {journal} {Journal of Applied Physics}\ }\textbf {\bibinfo {volume} {48}},\ \bibinfo {pages} {1412} (\bibinfo {year} {2008})}\BibitemShut {NoStop}%
\bibitem [{\citenamefont {Kaler}\ and\ \citenamefont {Jones}(1990)}]{kaler_dielectrophoretic_1990}%
  \BibitemOpen
  \bibfield  {author} {\bibinfo {author} {\bibfnamefont {K.~V.}\ \bibnamefont {Kaler}}\ and\ \bibinfo {author} {\bibfnamefont {T.~B.}\ \bibnamefont {Jones}},\ }\href {\doibase 10.1016/S0006-3495(90)82520-0} {\bibfield  {journal} {\bibinfo  {journal} {Biophysical Journal}\ }\textbf {\bibinfo {volume} {57}},\ \bibinfo {pages} {173} (\bibinfo {year} {1990})}\BibitemShut {NoStop}%
\bibitem [{\citenamefont {Bakewell}\ and\ \citenamefont {Morgan}(2006)}]{bakewell_dielectrophoresis_2006}%
  \BibitemOpen
  \bibfield  {author} {\bibinfo {author} {\bibfnamefont {D.}~\bibnamefont {Bakewell}}\ and\ \bibinfo {author} {\bibfnamefont {H.}~\bibnamefont {Morgan}},\ }\href {\doibase 10.1109/TNB.2005.864012} {\bibfield  {journal} {\bibinfo  {journal} {IEEE Transactions on NanoBioscience}\ }\textbf {\bibinfo {volume} {5}},\ \bibinfo {pages} {1} (\bibinfo {year} {2006})}\BibitemShut {NoStop}%
\bibitem [{\citenamefont {Cummings}\ and\ \citenamefont {Singh}(2003)}]{cummings_dielectrophoresis_2003}%
  \BibitemOpen
  \bibfield  {author} {\bibinfo {author} {\bibfnamefont {E.~B.}\ \bibnamefont {Cummings}}\ and\ \bibinfo {author} {\bibfnamefont {A.~K.}\ \bibnamefont {Singh}},\ }\href {\doibase 10.1021/ac0340612} {\bibfield  {journal} {\bibinfo  {journal} {Analytical Chemistry}\ }\textbf {\bibinfo {volume} {75}},\ \bibinfo {pages} {4724} (\bibinfo {year} {2003})}\BibitemShut {NoStop}%
\bibitem [{\citenamefont {Cummings}(2003)}]{cummings_streaming_2003}%
  \BibitemOpen
  \bibfield  {author} {\bibinfo {author} {\bibfnamefont {E.}~\bibnamefont {Cummings}},\ }\href {\doibase 10.1109/MEMB.2003.1266050} {\bibfield  {journal} {\bibinfo  {journal} {IEEE Engineering in Medicine and Biology Magazine}\ }\textbf {\bibinfo {volume} {22}},\ \bibinfo {pages} {75} (\bibinfo {year} {2003})}\BibitemShut {NoStop}%
\bibitem [{\citenamefont {Ding}\ \emph {et~al.}(2016)\citenamefont {Ding}, \citenamefont {M. Lawrence}, \citenamefont {V. Jones}, \citenamefont {G. Hogue},\ and\ \citenamefont {A. Hayes}}]{ding_concentration_2016}%
  \BibitemOpen
  \bibfield  {author} {\bibinfo {author} {\bibfnamefont {J.}~\bibnamefont {Ding}}, \bibinfo {author} {\bibfnamefont {R.}~\bibnamefont {M. Lawrence}}, \bibinfo {author} {\bibfnamefont {P.}~\bibnamefont {V. Jones}}, \bibinfo {author} {\bibfnamefont {B.}~\bibnamefont {G. Hogue}}, \ and\ \bibinfo {author} {\bibfnamefont {M.}~\bibnamefont {A. Hayes}},\ }\href {\doibase 10.1039/C5AN02430G} {\bibfield  {journal} {\bibinfo  {journal} {Analyst}\ }\textbf {\bibinfo {volume} {141}},\ \bibinfo {pages} {1997} (\bibinfo {year} {2016})}\BibitemShut {NoStop}%
\bibitem [{\citenamefont {Washizu}\ \emph {et~al.}(1994)\citenamefont {Washizu}, \citenamefont {Suzuki}, \citenamefont {Kurosawa}, \citenamefont {Nishizaka},\ and\ \citenamefont {Shinohara}}]{washizu_molecular_1994}%
  \BibitemOpen
  \bibfield  {author} {\bibinfo {author} {\bibfnamefont {M.}~\bibnamefont {Washizu}}, \bibinfo {author} {\bibfnamefont {S.}~\bibnamefont {Suzuki}}, \bibinfo {author} {\bibfnamefont {O.}~\bibnamefont {Kurosawa}}, \bibinfo {author} {\bibfnamefont {T.}~\bibnamefont {Nishizaka}}, \ and\ \bibinfo {author} {\bibfnamefont {T.}~\bibnamefont {Shinohara}},\ }\href {\doibase 10.1109/28.297897} {\bibfield  {journal} {\bibinfo  {journal} {IEEE Transactions on Industry Applications}\ }\textbf {\bibinfo {volume} {30}},\ \bibinfo {pages} {835} (\bibinfo {year} {1994})}\BibitemShut {NoStop}%
\bibitem [{\citenamefont {Kittel}(2004)}]{kittel_elementary_2004}%
  \BibitemOpen
  \bibfield  {author} {\bibinfo {author} {\bibfnamefont {C.}~\bibnamefont {Kittel}},\ }\href@noop {} {\emph {\bibinfo {title} {Elementary statistical physics}}},\ Dover books on physics\ (\bibinfo  {publisher} {Dover Publ},\ \bibinfo {address} {Mineola, NY},\ \bibinfo {year} {2004})\BibitemShut {NoStop}%
\bibitem [{\citenamefont {Bakewell}(2011)}]{bakewell_modelling_2011}%
  \BibitemOpen
  \bibfield  {author} {\bibinfo {author} {\bibfnamefont {D.~J.}\ \bibnamefont {Bakewell}},\ }\href {\doibase 10.1088/0022-3727/44/8/085501} {\bibfield  {journal} {\bibinfo  {journal} {Journal of Physics D: Applied Physics}\ }\textbf {\bibinfo {volume} {44}},\ \bibinfo {pages} {085501} (\bibinfo {year} {2011})}\BibitemShut {NoStop}%
\bibitem [{\citenamefont {Loucaides}\ \emph {et~al.}(2011)\citenamefont {Loucaides}, \citenamefont {Ramos},\ and\ \citenamefont {Georghiou}}]{loucaides_dielectrophoretic_2011}%
  \BibitemOpen
  \bibfield  {author} {\bibinfo {author} {\bibfnamefont {N.~G.}\ \bibnamefont {Loucaides}}, \bibinfo {author} {\bibfnamefont {A.}~\bibnamefont {Ramos}}, \ and\ \bibinfo {author} {\bibfnamefont {G.~E.}\ \bibnamefont {Georghiou}},\ }\href {\doibase 10.1016/j.elstat.2011.01.004} {\bibfield  {journal} {\bibinfo  {journal} {Journal of Electrostatics}\ }\textbf {\bibinfo {volume} {69}},\ \bibinfo {pages} {111} (\bibinfo {year} {2011})}\BibitemShut {NoStop}%
\bibitem [{\citenamefont {Castellanos}\ \emph {et~al.}(2003)\citenamefont {Castellanos}, \citenamefont {Ramos}, \citenamefont {González}, \citenamefont {Green},\ and\ \citenamefont {Morgan}}]{castellanos_electrohydrodynamics_2003}%
  \BibitemOpen
  \bibfield  {author} {\bibinfo {author} {\bibfnamefont {A.}~\bibnamefont {Castellanos}}, \bibinfo {author} {\bibfnamefont {A.}~\bibnamefont {Ramos}}, \bibinfo {author} {\bibfnamefont {A.}~\bibnamefont {González}}, \bibinfo {author} {\bibfnamefont {N.~G.}\ \bibnamefont {Green}}, \ and\ \bibinfo {author} {\bibfnamefont {H.}~\bibnamefont {Morgan}},\ }\href {\doibase 10.1088/0022-3727/36/20/023} {\bibfield  {journal} {\bibinfo  {journal} {Journal of Physics D: Applied Physics}\ }\textbf {\bibinfo {volume} {36}},\ \bibinfo {pages} {2584} (\bibinfo {year} {2003})}\BibitemShut {NoStop}%
\end{thebibliography}%

\appendix
\onecolumngrid
\pagebreak
\begin{center}
         \textbf{SUPPORTING INFORMATION}
     \end{center}

\begin{center}
         \textbf{Materials and Methods}
         \end{center}

\textbf{Chemicals.} (3-Aminopropyl)triethoxysilane (APTES, 99 $\%$), acetone, 2-propanol (IPA, $\geq$99.5 $\%$), ethanol ($\geq$99 $\%$), toluene (anhydrous, $\geq$99.8 $\%$), 50 nm, 100 nm, and 150 nm diameter gold nanoparticles (NPs, stabilized suspension in citrate buffer) were purchased from Sigma-Aldrich. 0.22 $\mu$m syringe filters with polytetrafluorethylen (PTFE) membrane were obtained from Whatman Anotop. 9 mm diameter imaging spacers were acquired from Grace Bio-Labs SecureSeal.

\textbf{Dielectrophoretic device fabrication.} Standard microelectronics techniques were employed for the dielectrophoretic device fabrication [S1, S2], and the detailed procedure is described elsewhere [S3]. Briefly, a 100 mm borosilicate wafer was cleaned in piranha solution and treated in high-frequency oxygen plasma. Standard photolithography procedures, including negative photoresist spin-coating, exposure, and development, were utilized to produce sawtooth microelectrode patterns on a borosilicate wafer. Next, the electron beam evaporation technique was used to deposit 5 nm Ti and 100 nm Au electrode material. Finally, the lift-off step and wafer dicing were conducted to produce 31 x 25 mm\textsuperscript{2} DEP chips.

\textbf{Numerical simulations.} The finite-elements simulations for the developed DEP device were performed using the AC/DC and Mathematics (classical partial differential equations, stabilized convection-diffusion equation) modules of COMSOL Multiphysics 6.1. The electric field intensity distribution was calculated near sawtooth metal electrodes based on the simulation of the potential $\phi$ distribution by solving the Laplace equation: $\nabla^2\phi=0$. The electric field thus was obtained as $E=-\nabla\phi$. Simulations were conducted for a unit cell of one gold sawtooth electrode pair attached to the 20 $\mu$m wide gold rectangle. The thickness of electrodes and the gap size between sawtooth pairs were 100 nm and 3 $\mu$m, respectively. Electrodes were located on 250.4 × 585.43 × 30 $\mu$m$^3$ (width × depth × height) SiO$_2$ substrate and immersed in the water medium of the same width and depth but with 120 $\mu$m height. One of the electrodes from the pair was grounded, while the boundary condition for the second was $\phi=V_{p-p}/2$, where $V_{p-p}=15$ V, which corresponds to the experimental value of the electric field. The frequency of the electric field was 3 MHz. Periodic boundaries were applied in $\pm x$ direction, and electric insulation conditions were for the remaining boundaries. The field was simulated in a water background, assuming a dielectric permittivity $\varepsilon_m$ = 78 and an electrical conductivity $\sigma_m$ of 16 $\mu$S/cm. Gold with $\sigma_{gold}$ = 456 kS/cm and $\varepsilon_{gold}$ = 6.9 was used for the electrode material. The minimum mesh size used for the discretization was 5 nm near the electrode apex.
The Au NPs concentration distribution was calculated in 2D in the plane of the DEP device surface by solving the modified particle-conservation equation (see Eq. (4) in the main text) to account for the particle steric effect [S4]. This modification is important to improve the convergence of the simulation results and obtain realistic particle concentrations. No flux boundary conditions were applied anywhere except the left and right boundaries of the simulation domain, for which Dirichlet boundary conditions were introduced, such that the particle volume fraction was kept constant, $c=0.001$. The initial volume fraction of Au nanoparticles in a simulation domain was also $c_0=0.001$. The minimum mesh size used for the discretization in this case was 5.6 nm in the gap between electrodes. The particle transport was simulated in a water background, assuming the same physical properties as at the electric field simulation step. The solution was obtained for a stationary condition such that the first term in Eq. (4) was set to zero.
The following parameters were taken for solving Eq. (4) in COMSOL: $R=$ 25 nm, 50 nm, 75 nm, $ D=k_BT/6\pi\eta R$ with $\eta=8.9\cdot10^4$ Pa$\cdot$s, $T = 300$ K, and $\mid \mathbf{E}(\mathbf{x}, \mathbf{y}, \mathbf{z}) \mid$ obtained by 3D COMSOL simulations.

\textbf{APTES functionalization of DEP device surface.} All DEP chips fabricated in this work were functionalized by APTES via a gas-phase deposition process. First, a DEP chip was thoroughly washed in acetone, IPA, and deionized water by placing it in an ultrasonic bath, followed by drying with N\textsubscript{2}. Next, the chip was treated in a high-frequency oxygen plasma (Tepla 300) at 1000 W for 5 min with 400 ml/min O\textsubscript{2} flow. After that, the chip was rewashed with ethanol, deionized water, and dried with N\textsubscript{2}. Immediately after cleaning and surface activation, the DEP chip was placed inside a glass beaker together with 1 mL of 99 $\%$ APTES, which was isolated in an opened vial to ensure efficient evaporation and prevent direct APTES contact with the chip. Subsequently, the glass beaker was tightly closed and placed in the oven (WTB Binder 7200) for 2 h at 70 \textdegree C. After the deposition, the chip was washed with a copious amount of toluene, ethanol, and water to remove unbound APTES residues and dried with N\textsubscript{2}. Finally, the APTES-modified DEP chip was placed in a clean glass Petri dish and annealed at 120 \textdegree C for 2 h to strengthen the chemical bonds between molecules and the surface.

\textbf{Characterization.} The reproducibility and morphology of the fabricated DEP device were characterized by field-emission SEM (Zeiss MERLIN) managed at 1 kV. The depletion region visualization was performed by an optical microscope (Leica DM8000) operated in a dark-field mode. Optical microscope (Nikon Optiphot 150) equipped by CCD camera (Chameleon 3 Color Camera, CM3-U3-50S5C-CS) and managed in dark-field mode with a 50  air objective (Nikon, NA = 0.55) was utilized to record all videos.

\textbf{DEP experiments and postprocessing.} All DEP experiments were performed for a colloidal solution of 25 nm, 50 nm, and 75 nm radius Au NPs redispersed in deionized water after double centrifugation for 30 min each at 1100$\times$g, 400$\times$g, and 300$\times$g, respectively. First, the device was energized by the function generator (GW Instek AFG-2125) by applying 15 V peak-to-peak AC voltage at the frequency of 3 MHz for various duration. Next, a drop of freshly prepared Au NPs aqueous solution (14 $\mu$L) was placed on top of the DEP device covered with an imaging spacer (Grace Bio-Labs SecureSeal) beforehand. A microscope coverslip was then placed on top of the spacer, avoiding a tight sealing to leave the possibility of removing unbound NPs after DEP, which was done by washing the device with copious amount of deionized water and drying with nitrogen. To record dark-field images of the DEP device after the experiment, the same optical microscope (Leica DM8000) was utilized. All acquired images were subsequently adjusted by image processing software (ImageJ) to enhance contrast and subtract the saturated optical signal and then analyzed to extract depletion region sizes.

\textbf{Quantitative analysis of Au nanoparticle concentrations.} We use Eq. (9) in the main text for analyzing the obtained concentration profiles and estimate the Au nanoparticle radius since the agreement between experiments and simulations is excellent. It is convenient to take the ratio of two surface concentrations, $c_{surf}(x_1,y_1)$ and $c_{surf}(x_2,y_2)$ to eliminate the need in defining the integration constant, \textit{A}, and absolute concentration values:
\begin{equation}
  \frac{c_{surf}(x_1,y_1)}{c_{surf}(x_2,y_2)} = \frac{Aexp\left[- \frac{\bigl \langle U_{DEP}(x_1, y_1)\bigr \rangle}{k_BT}\right]}{Aexp\left[- \frac{\bigl \langle U_{DEP}(x_2, y_2)\bigr \rangle}{k_BT}\right]} = exp \left[ \frac{R^3B}{k_BT} \left(\mid \mathbf{E} (x_2, y_2) \mid^2 - \mid \mathbf{E} (x_1, y_1) \mid^2 \right)\right], \tag{S1}
\end{equation}
where $c_{surf}(x_1,y_1)$ is the saturated high-value surface concentration obtained along the yellow line at $x_1$ = 120 $\mu$m from the electrode gap (see Figure 3c in the main text), $c_{surf}(x_1,y_1)$ is the concentration obtained along the same yellow line but in the depletion region, such that $c_{surf}(x_2,y_2)=c_{surf}(x_1,y_1)/2$, and $B=\pi \varepsilon_m \varepsilon_0 \operatorname{Re}\left[\frac{\varepsilon_p - \varepsilon_m}{\varepsilon_p+2\varepsilon_m}\right]$. Therefore, we can rewrite Eq. (S1) as:
\begin{equation}
  R = \left[ 
  \frac{k_BT \operatorname{ln}(2)}{B} \frac{1}{\left(\mid \mathbf{E} (x_2, y_2) \mid^2 - \mid \mathbf{E} (x_1, y_1) \mid^2 \right)}
  \right]. \tag{S2}
\end{equation}

It is seen from Eq. (S2) that the nanoparticle radius can be found from the difference of the electric field amplitude squared estimated at the corresponding surface coordinates \textit{x} and \textit{y}. We calculate the Au nanoparticle radii by taking the numerically simulated electric field values (see Figure S5) and assuming perfectly polarizable Au spheres $\left( \operatorname{Re}\left[\frac{\varepsilon_p - \varepsilon_m}{\varepsilon_p+2\varepsilon_m}\right] = 1\right)$, \textit{T} = 300 K, and $\varepsilon_m = 78$.

\begin{center}
         \textbf{Optimization of the experimental conditions}
         \end{center}

It is known that applying an electric field to a microfluidic system generates a series of effects on the fluid itself, including electrothermal (ET) flow, electroosmosis and natural convection, which can induce particle movement via the Stokes drag force [S5]. While this drag and other forces such as gravity and buoyancy could influence the measurement and have been discussed in the literature in great detail [S5, S6–S10], we focus here only on the two electrokinetic effects clearly observed in our experiments – DEP and ET effects – and how to keep them under control.

The ET effect also exists in highly inhomogeneous electric fields and induces two forces acting on a liquid: Coulomb and dielectric forces [S6, S11–S13]. The dominance of one or the other depends on the frequency range and determines the fluid flow direction, since both forces act in opposite directions. Hence, the fluid movement is minimal at the frequency where the transition between these two forces occurs. Besides, the fluid flow velocity determines the magnitude of the drag force for particles. In very high conductivity media, its magnitude may vastly exceed the DEP force, causing particles to follow the direction of the fluid flow. Such a particle behavior would refute our hypothesis that convection can be disregarded in Eq. (4) of the main text and we must ensure that the ET effect is suppressed in our DEP experiments.

The simplest way to suppress fluid flow is to reduce the liquid conductivity and its nonuniform heating to the lowest possible value. In this work, all DEP experiments were performed in water with a low conductivity of 16 $\mu$S/cm. Nevertheless, when we applied a sinusoidal signal with the amplitude of $V_{p-p}$ = 15 V and frequency of 500 kHz, we still observed rapid particle movement that is uncommon for DEP (see Video S1): particles circulated above the electrodes rather than being stably trapped in the gap between them. Detecting the depletion region in these conditions was impossible (Figure S6). Assuming that the ET effect is responsible for this particle movement, we gradually increased the electric field frequency up to 5 MHz, reaching the transition frequency between Coulomb and dielectric forces, where the fluid movement vanishes and DEP dominates (see Video S2). In this series of experiments, the optimum frequency of 3 MHz was determined, which is in agreement with the ET theory [S5, S6].

The second important factor that must be considered is the DEP electrode geometry. Indeed, along with generating a strong electric field to build well-discriminated depletion regions, the electrodes must also ensure sufficient space between adjacent DEP traps to prevent overlapping depletion regions. The effect of this intersection can be readily observed in the simulations of our DEP device when the distance between the gaps of two neighboring sawtooth electrode pairs is reduced from 250.4 $\mu$m to 83.5 $\mu$m (see Figure S7). In this case, the concentration profiles near each electrode pairs merge, which blurs their boundaries. Therefore, one must carefully optimize the electrodes geometry and experimental conditions such that there is a spacing of at least two times the depletion region.

\begin{center}
         \textbf{References}
         \end{center}

[S1] Madou, M. J. \textit{Manufacturing Techniques for Microfabrication and Nanotechnology}; CRC press, \textbf{2011}; Vol.2.

[S2] Abasahl, B.; Santschi, C.; Raziman, T. V.; Martin, O. J. F. \textit{Nanotechnology} \textbf{2021}, \textit{32}, No. 475202.

[S3] Zavatski, S.; Bandarenka, H.; Martin, O. J. F. \textit{Anal. Chem.} \textbf{2023}, \textit{95}, 2958–2966.

[S4] Loucaides, N. G.; Ramos, A.; Georghiou, G. E. \textit{Journal of Electrostatics} \textbf{2011}, \textit{69} (2), 111–118.

[S5] Castellanos, A.; Ramos, A.; Gonzalez, A.; Green, N. G.; Morgan, H.; \textit{Journal of Physics D} \textbf{2003}, \textit{36} (20), 2584.

[S6] Ramos, A.; Morgan, H.; Green, N. G.; Castellanos, A. \textit{Journal of Physics D: Applied Physics} \textbf{1998}, \textit{31} (18), 2338–2353.

[S7] Green, N. G.; Morgan, H.; Milner, J. J. \textit{Journal of Biochemical and Biophysical Methods} \textbf{1997}, \textit{35} (2), 89–102.

[S8] Green, N. G.; Ramos, A.; Morgan, H. \textit{Journal of Physics D: Applied Physics} \textbf{2000}, \textit{33} (6), 632–641.

[S9] Green, N. G.; Morgan, H. \textit{Journal of Physical Chemistry B} \textbf{1999}, \textit{103} (1), 41–50.

[S10] Green, N. G.; Ramos, A.; González, A.; Castellanos, A.; Morgan, H. \textit{Journal of Physics D: Applied Physics} \textbf{2000}, \textit{33} (2).

[S11] Green, N. G.; Ramos, A.; González, A.; Castellanos, A.; Morgan, H. \textit{Journal of Electrostatics} \textbf{2001}, \textit{53} (2), 71–87. 

[S12] Salari, A.; Navi, M.; Lijnse, T.; Dalton, C. \textit{Micromachines} \textbf{2019}, \textit{10} (11), 1–27. 

[S13] Sun, H.; Ren, Y.; Hou, L.; Tao, Y.; Liu, W.; Jiang, T.; Jiang, H. \textit{Analytical Chemistry} \textbf{2019}, \textit{91} (9), 5729–5738.

\clearpage
\onecolumngrid
\begin{center}
         \textbf{SUPPORTING FIGURES}
     \end{center}
\setcounter{figure}{0}
\makeatletter 
\renewcommand{\thefigure}{S\@arabic\c@figure}
\makeatother
\begin{figure*}[!htpb]
  \centering
  \includegraphics[width=0.9\linewidth]{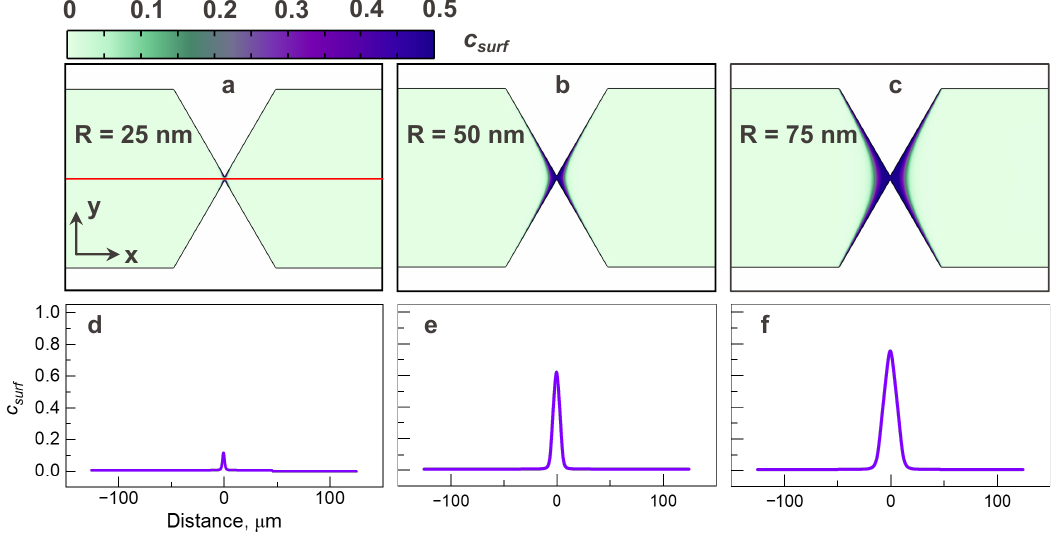}
  \caption{2D simulation results of the concentration distributions right above the DEP device surface for (a, d) 25 nm, (b, e) 50 nm, and (c, f) 75 nm radius Au nanoparticles after applying a sinusoidal electric signal with 15 V peak–to–peak voltage and 3 MHz frequency. The concentration distribution profiles in (d–f) were calculated along the red line in (a) crossing the middle of the gap between adjacent electrode pairs.}
  \label{fgr:figS1}
\end{figure*}
\begin{figure*}[!htpb]
  \centering
  \includegraphics[width=0.9\linewidth]{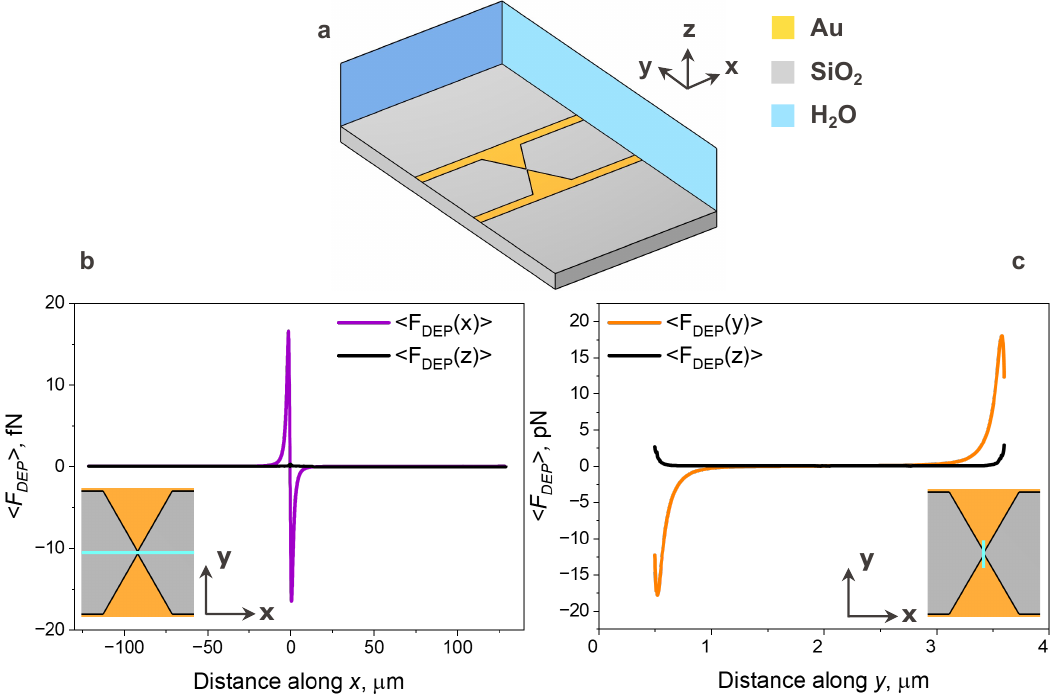}
  \caption{(a) Schematic geometry of the DEP device used to simulate (b) \textit{x}- and \textit{z}-components and (c) \textit{y}- and \textit{z}-components of the DEP force, $\bigl \langle\mathbf{F_{DEP}}(x)\bigr \rangle$, $\bigl \langle\mathbf{F_{DEP}}(y)\bigr \rangle$, and $\bigl \langle\mathbf{F_{DEP}}(z)\bigr \rangle$ along the cyan lines shown in the corresponding insets.}
  \label{fgr:figS2}
\end{figure*}
\begin{figure*}[!htpb]
  \centering
  \includegraphics[width=0.9\linewidth]{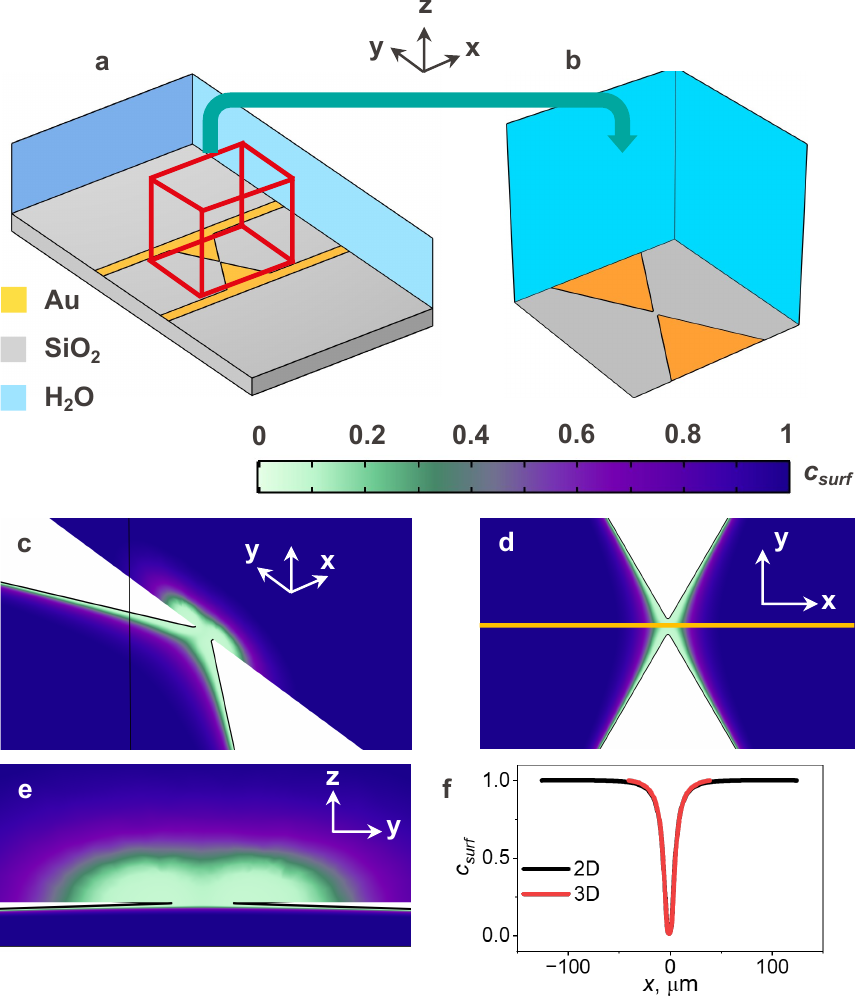}
  \caption{Schematic geometry of the DEP device used to simulate electric field components $\mathbf{E}(x)$, $\mathbf{E}(y)$, and $\mathbf{E}(z)$, which were transferred to (b) a reduced 3D simulation domain with width × depth × height of 80 × 80 × 80 $\mu$m for simulating concentration distribution of Au nanoparticles during DEP. (c) 3D, (d) top and (e) cross–sectional view of the simulation results of the concentration distributions for 25 nm radius Au nanoparticles after applying a sinusoidal electric signal with 15 V peak–to–peak voltage and 3 MHz frequency. (f) The comparison of concentration distribution profiles in 2D and 3D were calculated along the yellow line in (d) crossing the middle of the gap between adjacent electrode pairs.}
  \label{fgr:figS3}
\end{figure*}
\begin{figure*}[!htpb]
  \centering
  \includegraphics[width=\linewidth]{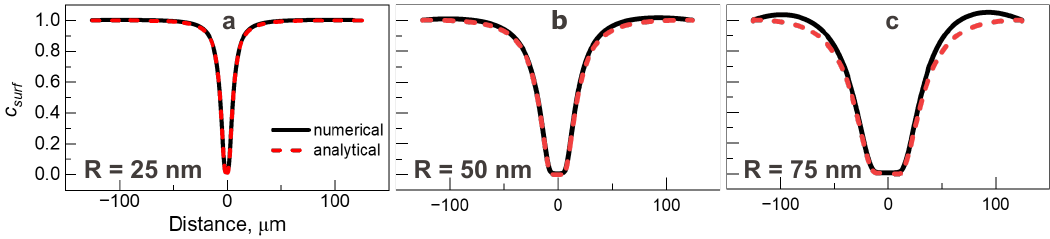}
  \caption{Comparison of concentration profiles of (a) 25 nm, (b) 50 nm, and (c) 75 nm Au nanoparticles obtained by numerical simulations and calculations using Eq. (9) indicated in the main text.}
  \label{fgr:figS4}
\end{figure*}
\begin{figure*}[!htpb]
  \centering
  \includegraphics[width=0.7\linewidth]{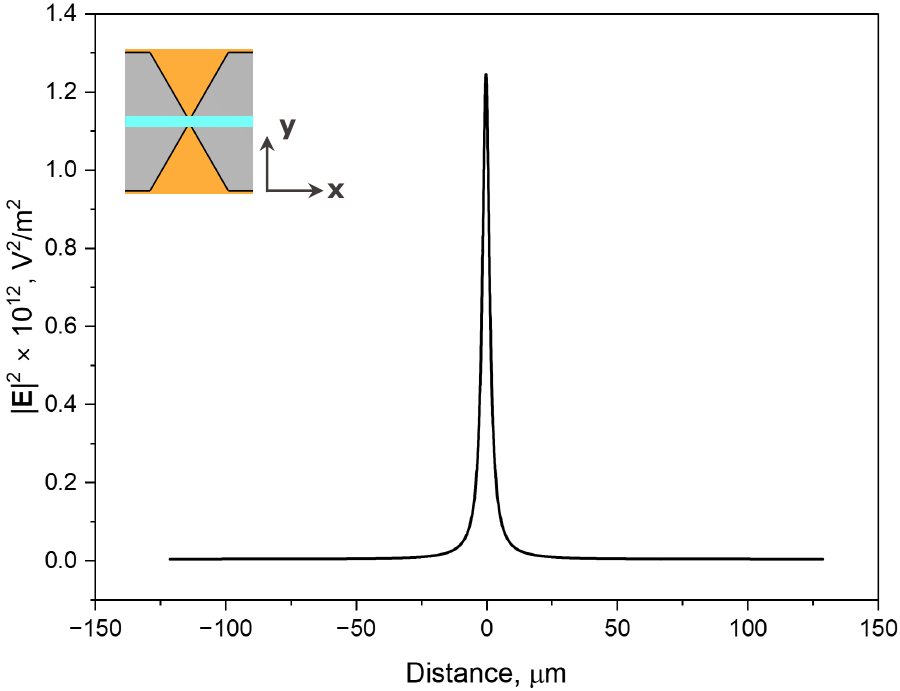}
  \caption{Magnitude of the electric field squared simulated for sawtooth electrode pairs and plotted along the cyan line shown in the inset.}
  \label{fgr:figS5}
\end{figure*}
\begin{figure*}[!htpb]
  \centering\includegraphics{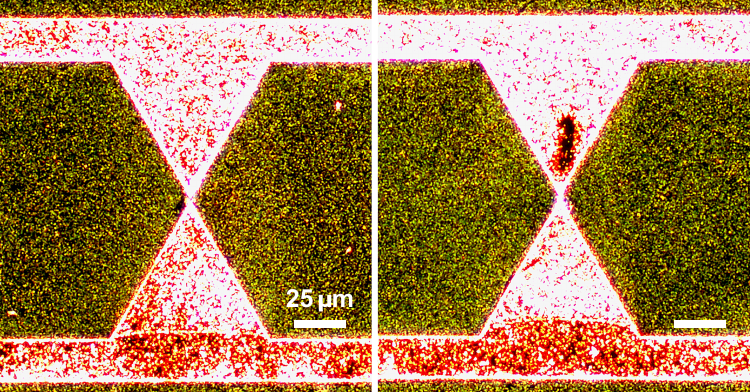}
  \caption{Magnitude of the electric field squared simulated for sawtooth electrode pairs and plotted along the cyan line shown in the inset.}
  \label{fgr:figS6}
\end{figure*}
\begin{figure*}[!htpb]
  \centering
  \includegraphics[width=\linewidth]{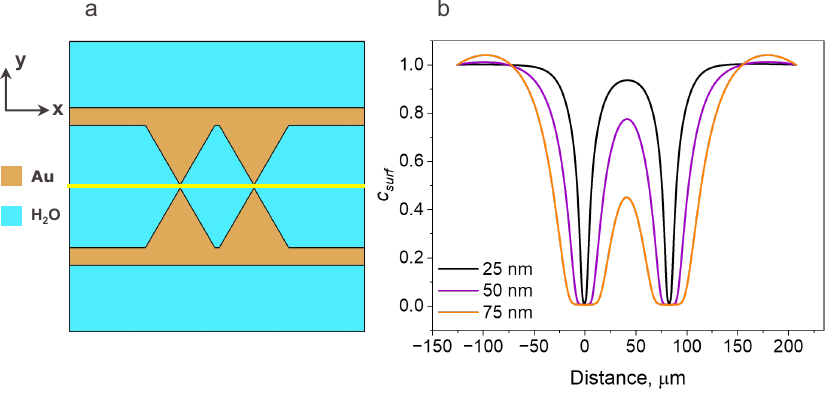}
  \caption{(a) Schematic view of sawtooth electrode pairs with the reduced from 250.4 $\mu$m to 83.5 $\mu$m intergap distance and (b) simulated concentration profiles plotted along the yellow line in (a).}
\end{figure*}

\end{document}